\documentclass[sigconf]{aamas} 
\usepackage{balance} % for balancing columns on the final page
\usepackage{graphicx}
\usepackage{algorithm}
\usepackage{amsmath}
\usepackage{amsthm}
\usepackage{subcaption}
\usepackage{thmtools,thm-restate}
\usepackage{color}
\usepackage[noend]{algpseudocode}
\usepackage{enumerate}
\usepackage{cases}
\usepackage{multicol}
\usepackage{multirow}
\usepackage{enumitem}
\usepackage{bm}
\usepackage{import}

\setcopyright{ifaamas}
\acmConference[AAMAS '22]{Proc.\@ of the 21st International Conference
on Autonomous Agents and Multiagent Systems (AAMAS 2022)}{May 9--13, 2022}
{Online}{P.~Faliszewski, V.~Mascardi, C.~Pelachaud,
M.E.~Taylor (eds.)}
\copyrightyear{2022}
\acmYear{2022}
\acmDOI{}
\acmPrice{}
\acmISBN{}

\acmSubmissionID{190}

\title{A Distributed Differentially Private Algorithm for Resource Allocation in Unboundedly Large Settings}

\author{Panayiotis Danassis}
\affiliation{
  \institution{\'Ecole Polytechnique F\'ed\'erale de Lausanne (EPFL)}
  \city{Lausanne}
  \country{Switzerland}}
\email{panayiotis.danassis@alumni.epfl.ch}

\author{Aleksei Triastcyn}
\affiliation{
  \institution{\'Ecole Polytechnique F\'ed\'erale de Lausanne (EPFL)}
  \city{Lausanne}
  \country{Switzerland}}
\email{aleksey.tryastsyn@alumni.epfl.ch}

\author{Boi Faltings}
\affiliation{
  \institution{\'Ecole Polytechnique F\'ed\'erale de Lausanne (EPFL)}
  \city{Lausanne}
  \country{Switzerland}}
\email{boi.faltings@epfl.ch}

\begin{abstract}
We introduce a \emph{practical} and \emph{scalable} algorithm (PALMA) for solving one of the fundamental problems of multi-agent systems -- finding matches and allocations -- in \emph{unboundedly large} settings (e.g., resource allocation in urban environments, mobility-on-demand systems, etc.), while providing \emph{strong worst-case privacy} guarantees. PALMA is decentralized, runs on-device, requires no inter-agent communication, and converges in constant time under reasonable assumptions. We evaluate PALMA in a mobility-on-demand and a paper assignment scenario, using \emph{real data} in both, and demonstrate that it provides a strong level of privacy ($\varepsilon \leq 1$ and median as low as $\varepsilon = 0.5$ across agents) and high-quality matchings (up to $86\%$ of the non-private optimal, outperforming even the privacy-preserving centralized maximum-weight matching baseline).
\end{abstract}

\newtheorem{definition}{Definition}

\makeatletter
\newcommand\footnoteref[1]{\protected@xdef\@thefnmark{\ref{#1}}\@footnotemark}
\makeatother

\begin{document}

%%% The following commands remove the headers in your paper. For final 
%%% papers, these will be inserted during the pagination process.

\pagestyle{fancy}
\fancyhead{}

%%% The next command prints the information defined in the preamble.

\maketitle 

%%%%%%%%%%%%%%%%%%%%%%%%%%%%%%%%%%%%%%%%%%%%%%%%%%%%%%%%%%%%%%%%%%%%%%%%

\section{Introduction} \label{sec:Introduction}

One of the fundamental problems in multi-agent systems is finding an optimal allocation, i.e., solving a maximum-weight matching (MWM) problem. A wide range of applications -- spanning from mobility-on-demand systems and ridesharing~\cite{danassis2019putting} to kidney exchange~\cite{roth2005pairwise} -- can be formulated and solved as a weighted matching problem. Real-world matching problems pose three significant challenges: (i) they may occur in \emph{unboundedly large} settings (e.g., resource allocation in urban environments), (ii) they are \emph{distributed} and \emph{information-restrictive} (agents have partial observability and inter-agent communication might not be available~\cite{AAAI10-adhoc}), and finally, (iii) individuals have to \emph{reveal their preferences} in order to get a high-quality match, which brings forth significant privacy risks. In this work, we propose \emph{PALMA} (Privacy-preserving ALtruistic MAtching), a matching algorithm designed to tackle \emph{all} of the aforementioned challenges.

PALMA is a privacy-preserving adaptation of ALMA~\cite{ijcai201931,ijcai2021-18,danassis2022Scalable}; a recently proposed algorithm for real-world, large-scale applications that solves the first two challenges. As such, it is \emph{decentralized}, requires \emph{no communication} between the participants, and converges in \emph{constant} (to the total problem size) time -- in the realistic case where each agent is interested in a (fixed size) subset of the total resources.

The third challenge requires \emph{protecting the utility functions} of the agents. In recent years, Differential Privacy (DP)~\cite{10.1007/11787006_1} (and its variants) has emerged as the de facto standard for protecting the privacy of individuals. Informally, a DP algorithm ensures indistinguishability on the output distributions for any neighboring inputs. We have designed a defense mechanism for PALMA based on the idea of randomized response~\cite{warner1965randomized} -- which involves adding controlled randomness -- that results in indistinguishability under Local DP~\cite{dwork2014algorithmic}.

One final challenge arises when it comes to \emph{large-scale} multi-agent systems with a \emph{diverse} set of agents, as it is hard to achieve a meaningful privacy guarantee -- in a practical way -- using standard (L)DP if the problem has a large output space (e.g., matches, and allocations)~\cite{7932877,10.1145/2591796.2591826}. Conventional (L)DP mechanisms often require adding a lot of random noise to achieve a meaningful privacy guarantee, which in turn leads to a pronounced drop in the solution quality. More often than not, this is not due to the inherent difficulty of the problem at hand, but rather due to the generality of the DP definition. Not only does DP consider a very broad class of adversaries, it also protects all users -- independent of their characteristics -- by the same guarantee. While this property is being praised as one of the strongest arguments in favor of DP, it can be completely redundant in many real-world applications for three key reasons: (i) users might be willing to disclose less-sensitive information (e.g., city of residence, but not exact location), (ii) the attacker might already know coarser-grained information because it is likely public or easily available and, thus, does not need to be hidden (e.g., city of residence in a mobility-on-demand system, or reviewer expertise in a paper assignment problem), and (iii) domain characteristics might exclude a subset of solutions (e.g., a taxi in Manhattan will not be assigned to serve a request in Brooklyn, and an expert on auctions would not be assigned to review a robotics paper, thus, there is no need for indistinguishably between taxis in different boroughs or reviewers on different fields).

To solve this challenge, we motivate and develop a `context-aware' privacy definition (\emph{Piecewise Local Differential Privacy} -- PLDP), which takes into account the `distance' between the images of two utility functions. The level of protection depends on that distance; agents with utility functions that have images close in distance to each other would be indistinguishable from the attacker's point of view. 

% The definition is inspired by existing work on `data-aware'~\cite{triastcyn2020bayesian,triastcyn2020data} and distance-based generalisations of DP~\cite{chatzikokolakis2013broadening,andres2013geo}.

% \note{Despite notable similarities with the latter, there is an important difference. Instead of being centered around the agent, privacy-protected regions are predefined by some function $\varphi(\cdot)$ that partitions the space of possible outcomes $\mathcal{D}$ into subspaces $\{ \mathcal{D}_i \}$. Importantly, this allows us to use \emph{tighter composition theorems} developed for the conventional DP, which gives \emph{significant advantage} in real-world settings by \emph{reducing the growth} of $\varepsilon$ over the iterations.} We provide more details in Section~\ref{sec:piecewise_local_privacy}.
% % Finally, we combine PLDP with ALMA to create a decentralized, privacy-preserving algorithm (PALMA) for large-scale weighted matching problems.

\subsection{Our Contributions} \label{sec:Our Contributions}

\noindent
\textbf{(1) We propose \emph{PALMA}, the \emph{first} practical and scalable privacy preserving algorithm} for weighted matching in \emph{unboundedly large} settings with thousands of agents (e.g., resource allocation in urban environments, intelligent infrastructure, IoT devices, etc.).
%  % which combines ALMA~\cite{ijcai201931}, PLDP, and a privacy accounting method for iterative algorithms.

\noindent
\textbf{(2) We introduce \emph{Piecewise Local Differential Privacy} (PLDP)}, a variant of differential privacy designed to protect the utility function in multi-agent applications. PLDP enables significant improvements in solutions quality and strong theoretical privacy guarantees, while being applicable in \emph{real-world, unboundedly large settings}.
% designed to protect the utility function in multi-agent applications where the attacker possesses additional information on the characteristics of the utility space.

% \smallskip\noindent
% \textbf{(2) We design a novel privacy \emph{accounting} method for \emph{iterative algorithms}}. It generalizes and encapsulates several previously known and widely used methods, such as the moments accountant, thereby providing a clean, unified framework for accounting differential privacy in iterative algorithms. 

\noindent
\textbf{(3) We evaluate PALMA in a mobility-on-demand and a paper assignment scenario, using \emph{real data}}. PALMA is able to provide a high degree of privacy,  $\varepsilon \leq 1$ and a median value as low as $0.5$ across agents  for $\delta = 10^{-5}$, and matchings of high quality (up to $86\%$ of the non-private optimal).

% As a bonus contribution, we extend the existing tight adaptive accounting for Gaussian subsampled mechanisms to general subsampled mechanisms.
% following the same proving technique as in~\cite{triastcyn2020bayesian}.

The decentralized algorithm and corresponding privacy definition allows PALMA to adapt the noise added for obfuscation to the privacy budget of each agent. This achieves significantly better performance than the centralized Hungarian algorithm with the fixed obfuscation required to achieve the same privacy guarantees with an untrusted server. 

% \emph{Ultimately, we are the first to develop a practical and scalable framework for weighted matching and resource allocation in general, unboundedly large, multi-agent systems}.

\subsection{Related Work} \label{sec:Related Work}

Finding a maximum-weight matching is one of the best-studied combinatorial optimization problems~\cite{su2015algorithms,lovasz2009matching}. Yet, while the problem has been `solved' from an algorithmic perspective -- having both centralized and decentralized polynomial algorithms -- it is not so from the perspective of multi-agent systems, for three key reasons: (i) \emph{complexity}, (ii) \emph{communication}, and (iii) \emph{privacy}.
% There is a plethora of centralized polynomial time algorithms (e.g., Hungarian~\cite{kuhn1955hungarian}, blossom~\cite{edmonds1965maximum}). In real-world problems, a centralized coordinator is not always available, and if so, it has to \emph{know the utilities} of all the participants which is often not feasible and poses significant privacy risks. Decentralized algorithms (e.g.,~\cite{7991447,zavlanos2008distributed}) require polynomial computational time and polynomial number of messages. Thus,

The proliferation of intelligent systems will give rise to large-scale, multi-agent based technologies. Algorithms for maximum-weight matching, whether centralized or distributed, have runtime that increases with the total problem size, even in the realistic case where agents are interested in a small number of resources. Thus, they can only handle problems of bounded size. Moreover, they require a significant amount of inter-agent communication. Yet, communication might not always be an option~\cite{AAAI10-adhoc}, and sharing utilities, plans, and preferences creates high overhead. ALMA on the other hand achieves \emph{constant} in the total problem size running time -- under reasonable assumptions -- while requiring no message exchange (i.e., no communication network) between the participating agents~\cite{ijcai201931}. The proposed approach, PALMA, \emph{preserves} the aforementioned two properties of ALMA, thus, dealing with the first two of the posed challenges.

Differential Privacy (DP)~\cite{dwork2006,10.1007/11787006_1,dwork2006our,dwork2006calibrating} has emerged as the de facto standard for protecting the privacy of individuals (see Appendix \ref{supp: Differential Privacy Definition} for the definition of DP, along with intuitive examples)\footnote{For a more comprehensive overview, we refer the reader to~\cite{triastcyn2020data,dwork2014algorithmic}.}. Informally, DP captures the increased risk to an individual's privacy incurred by his participation. A variation of differential privacy, especially useful in our context, given the decentralized nature of PALMA, is Local Differential Privacy (LDP)~\cite{dwork2014algorithmic}. LDP is a generalization of DP that provides a bound on the outcome probabilities for any pair of individual agents rather than populations differing on a single agent. Intuitively, it means that one cannot hide in the crowd. Another strength of LDP is that it does not use a centralized model to add noise---individuals sanitize their data themselves---providing privacy protection against a malicious data curator. As a result, LDP requires adding even more random noise to achieve a meaningful bound, which would result in the decline of the solution quality. In fact, it is impossible to have both meaningful social welfare and privacy guarantees in matching problems under (L)DP~\cite{10.1145/2591796.2591826}. (L)DP ignores specifics of AI applications, such as a focus on a given task or a particular data distribution.

Our work is inspired by the literature on `data-aware' privacy notions~\cite{triastcyn2020bayesian,triastcyn2020data} and distance-based generalisations of DP~\cite{chatzikokolakis2013broadening,andres2013geo}. As a matter of fact, there are works that utilize such distance-based notions to solve a weighted matching problem in specific domains (e.g.,~\cite{prorok2017privacy,10.5555/3237383.3237910}). Yet, these are centralised approaches and thus face a (computation and communication) complexity barrier (refer back to the aforementioned challenges (i) and (ii) of real-world matching problems). ALMA can also be combined with other existing notions of privacy (e.g., LDP or geo-indistinguishability~\cite{andres2013geo}), yet the solution quality is inferior compared to the proposed, carefully crafted (with ALMA in mind) noise, as we demonstrate in our evaluation.

\section{Piecewise Local Differential Privacy (PLDP)}
\label{sec:piecewise_local_privacy}
% In this section, we provide a detailed description of our privacy model, named \emph{Piecewise Local Differential Privacy (PLDP)}.

Inspired by the notions of Bayesian DP \cite{DBLP:conf/bigdataconf/TriastcynF19} -- which is based on the observation that machine learning models are designed and tuned for a particular data distribution which is also often available to the attacker -- and metric-based DP \cite{chatzikokolakis2013broadening} and geo-indistinguishability \cite{andres2013geo} -- where indistinguishability depends on an arbitrary notion of distance -- we propose a new privacy model, namely Piecewise Local Differential Privacy (PLDP). PLDP takes into account the `distance' between the images of two utility functions, and the level of protection depends on that distance. The rationale is that instead of guaranteeing local privacy in the entire domain of agents, which can be quite difficult and would result in low quality solutions due to excessive noise, we focus on indistinguishability of agents with similar preferences.
% -- which is based on the observation that machine learning models are designed and tuned for a \emph{particular data distribution} (e.g., an MRI dataset is very unlikely to contain a picture of a car) and such prior distribution of data is also often \emph{available to the attacker} --

% \subsection{Definition}
% \label{sec:pldp_definition}
Let $\mathcal{M} : \mathcal{D} \rightarrow \mathcal{A}$ be a randomized function with domain $\mathcal{D}$ and range $\mathcal{A}$. In the context of matching problems in multi-agent systems, $\mathcal{D}$ is the space of utility functions and $\mathcal{A}$ is the action space.
% In the context of PALMA specifically, an action is either an attempt to acquire a certain resource, or a back-off from a previously contested resource, as will explained in the following section.

% (possibly overlapping)
\begin{definition}
\label{def:pldp}
Let $\varphi(\cdot)$ be a set function that fragments $\mathcal{D}$ into a collection of subsets $\{ \mathcal{D}_i \}$. Then, a randomized algorithm $\mathcal{M}: \mathcal{D} \rightarrow \mathcal{A}$ satisfies $(\varepsilon, \delta, \varphi)$-piecewise local privacy if for any two inputs $x, x' \in \mathcal{D}_i,~\forall i$, and for any set of outcomes $\mathcal{S} \subset \mathcal{A}$ it holds:
\begin{align*}
    \Pr\left[\mathcal{M}(x) \in \mathcal{S} ~|~ x \in \mathcal{D}_i \right] \leq e^\varepsilon \Pr\left[\mathcal{M}(x') \in \mathcal{S} ~|~ x' \in \mathcal{D}_i  \right] + \delta.
\end{align*}
\end{definition}

\subsection{Motivation} \label{PLDP Motivation}

Consider a mobility-on-demand (MoD) application (e.g., ridesharing). A MoD company can operate across multiple cities, countries, or even continents. If a MoD provider employs traditional DP (e.g., LDP) to protect all users (independently of their characteristics) with the same guarantee, the achieved social welfare will be as good as a \emph{random solution}\footnote{The solution that results of picking edges randomly in a fully connected bipartite graph containing all agents and resources.} in large-scale environments. This is because the \emph{support} of any agent has to include \emph{every resource} (otherwise an adversary could distinguish between agents), i.e., a request in Manhattan might be paired with a taxi in Brooklyn. Moreover, it is reasonable to assume an informed attacker (e.g., one that knows the city of residence), and users may be willing to reveal approximate location information (it is most likely acceptable to disclose the fact that an individual is in Manhattan, however disclosing the exact location is undesirable). Similarly, in a paper assignment problem (reviewers to manuscripts), ensuring indistinguishably between an expert on Markets \& Auctions, and one on Robotics might be futile, especially if the attacker possesses additional information (e.g., the tracks of the papers) that would exclude infeasible matches.

The rationale behind PLDP is the following. Instead of guaranteeing local privacy in the entire domain of agents, which may be quite difficult, we focus on indistinguishability of agents with \emph{similar preferences}. We fragment the space of utilities into regions and guarantee privacy within these regions but not between them.

A useful real-world analogy is ZIP codes. Assume we would like to release some location statistic with PLDP and we choose $\varphi$ such that the initial location space is mapped into ZIP codes. Then, $(\varepsilon, \delta, \varphi)$-PLDP guarantee would certify that the reported statistic is $(\varepsilon, \delta)$-locally private within every ZIP code. However, it would not tell us anything about privacy of the reported statistic outside the given ZIP code. In other words, while an agent can be distinguished from agents outside his zip code, he is still indistinguishable from \emph{all} agents inside his ZIP code.

\subsection{Privacy Properties}

Note that PLDP is a straightforward relaxation of local privacy and all the properties of LDP are satisfied within sub-domains $\mathcal{D}_i$. In order to see that this is true, it is sufficient to consider the following. Once the space $\mathcal{D}$ has been partitioned, the PLDP definition is equivalent to the LDP definition within each sub-space $\mathcal{D}_i$. Hence, basic properties of (L)DP, such as \emph{composition}, \emph{post-processing}, and \emph{group privacy}, as well as several instances of \emph{advanced composition}~\cite{dwork2014algorithmic,abadi2016deep}, will also hold for any pair $x,x'$ from a given $\mathcal{D}_i$, as long as these points do not dynamically change sub-domains between applications of the privacy mechanism. The latter condition is satisfied in all considered scenarios: every new matching routine starts with a fresh set of agents with random identifiers, and agents do not change their utilities during the matching process.

\subsection{Advantages of PLDP (vs. Distance-based Generalisations of DP)} \label{Advantages of PLDP}

PLDP closely resembles another well-known privacy notion, geo-indistinguishability~\cite{andres2013geo}, which is based on a generalization of DP~\cite{chatzikokolakis2013broadening}. Nonetheless, there is a notable distinction. To put it in terms of the definition above, in geo-indistinguishability, the region within which privacy is protected is centered at $x$. In our definition, these regions are predefined by $\varphi$. As a downside, our privacy guarantee is limited to the given region rather than fading gradually with increasing region radius. However, there is also a crucial upside to this subtle difference in real-world applications due to composition properties. To the best of our knowledge, in spite of conveniently adopting the use of distances between inputs to adjust levels of privacy guarantees, geo-indistinguishability has only been proven to satisfy basic composition. As a result, $\varepsilon$ grows linearly with the number of privacy mechanism invocations. It is not sufficiently tight for iterative AI and ML applications, which typically require a lot of repetitive applications of privacy mechanisms~\cite{abadi2016deep}. On the other hand, PLDP allows to use \emph{tighter composition theorems} developed for the conventional DP, reducing the growth of $\varepsilon$ from linear w.r.t. the total number of algorithm iterations $T$ to $\mathcal{O}(\sqrt{T})$~\cite{abadi2016deep}.

A second advantage of PLDP is that, contrary to geo- indistinguishability, it does not require a metric space (i.e., a natural ordering). As an example, this makes PLDP easier to apply in settings like our paper assignment application where each agent/resource is represented by a $25$-dimensional binary label (see Appendix \ref{supp: Paper Assignment}). In this example, there is ordering in each dimension, but not across them.

 % is that PLDP is easy to use whether or not there is a natural ordering of attributes, while geo-indistinguishability requires a distance metric. 
 % i.e., we have more applicability.

% A second advantage for PLDP arises in discrete settings. Geo-indistinguishability requires to discretize the metric space by partitioning it into cells. Then, the probability of a cell incorporates the probability mass obtained by the integration of the probability density function over the cell~\cite{8429310,andres2013geo}. I.e., additional noise is needed to compensate the effect of discretization, which results in a degradation of the privacy guarantee. \note{Contrary to that...}

\section{PALMA: A Privacy-Preserving Weighted Matching Algorithm} \label{sec: PALMA}

% In this section we introduce \emph{PALMA} (Privacy-preserving ALtruistic MAtching), a privacy-preserving adaptation of ALMA~\cite{ijcai201931}. We start by describing the problem of finding a maximum-weight matching. For simplicity, we will focus on bipartite graphs (i.e., the assignment problem), but (P)ALMA can be applied in general graphs as well (see~\cite{danassis2019putting} for an example application of ALMA in general graphs). Finally, we describe the employed privacy mechanisms, and the privacy accounting method.

\subsection{The Assignment Problem} \label{The Assignment Problem}

The assignment problem refers to finding a maximum-weight matching in a weighted bipartite graph\footnote{ALMA (and thus PALMA) can be applied in general graphs as well (see~\cite{danassis2019putting}).}, $\mathcal{G} = \left\{ \mathcal{N} \cup \mathcal{R}, \mathcal{E} \right\}$. In the studied scenario, $\mathcal{N} = \{1, \dots, N\}$ agents compete to acquire $\mathcal{R} = \{1, \dots, R\}$ resources. The weight of an edge $(n, r) \in \mathcal{E}$ represents the utility ($u_n(r) \in [0, 1]$) agent $n$ receives by acquiring resource $r$. Each agent can acquire at most one resource, and each resource can be assigned to at most one agent. The goal is to maximize the sum of utilities.

% , i.e., $\max_{\mathbf{x} \geq 0} \sum_{(n,r) \in \mathcal{E}} u_n(r) x_{n, r}$ -- where the variable $x_{n, r}$ is 1 if the edge is contained in the matching and 0 otherwise, and $\mathbf{x} = (x_{1, 1}, \dots, x_{N, R})$ -- subject to $\sum_{r | (n,r) \in \mathcal{E}} x_{n, r} = 1, \forall n \in \mathcal{N}$, and $\sum_{n | (n,r) \in \mathcal{E}} x_{n, r} = 1, \forall r \in \mathcal{R}$.

For simplicity, in the rest of the paper we assume $N = R$. This is \emph{not required} by PALMA (or ALMA). If $R > N$ some resources will remain free, while if $N > R$ some agents will fail to acquire a resource (convergence in the latter case implies that the state of the agent does not change, see~\cite{ijcai201931}).

\subsection{Learning Rule} \label{sec: Learning Rule}

We assume each agent is interested in (potentially) a subset of the total resources $\mathcal{Q}^n \subseteq \mathcal{R}$. Let $\mathcal{A} = \{Y, A_{r_1}, \dots, A_{r_{Q^n}}\}$ denote the set of actions, where $Y$ refers to yielding, and $A_r$ refers to accessing resource $r$. Let $g$ denote the agent's strategy. PALMA is run \emph{independently and in parallel by all the agents}. Each agent converges to a resource through repeated trials, specifically:

As long as an agent has not acquired a resource yet, at every time-step, there are two possible scenarios: If $g = A_r$ (strategy points to resource $r$), then agent $n$ attempts to acquire that resource. If there is a collision\footnote{\label{footnote: no communication}We assume that agents can observe feedback from their environment to inform collisions and detect free resources (e.g., by the use of sensors, or by a single bit feedback from the resource).}, the colliding parties back-off with some probability, $P_B^n(\cdot)$. Otherwise, if $g = Y$, the agent chooses a resource $r$ for monitoring according to probability , $P_S^n(\cdot)$. If the resource is free, he sets $g \leftarrow A_r$. The pseudo-code can be found in Alg. \ref{algo: palma}.

\subsubsection{\textbf{Resource Selection Distribution}} \label{sec: Resource Selection Distribution}

In the original ALMA, each agent sorts the resources in decreasing order of utility ($r_1, \dots, r_{R}$). Then, he moves in a sequential manner, starting from the most preferred resource ($r_1$), and moving down the list until he acquires one. This method of resource selection results in the highest social welfare, but it is impossible to guarantee privacy due to the deterministic nature of the selection process. On the other end of the spectrum, we can select a resource in a weighted at random fashion, where resource $r_i$ is selected with probability $\frac{u_n(r_i)}{\sum_{r \in \mathcal{R}} u_n(r)}$. This method provides high degree of privacy, but can result in low social welfare. To elaborate the latter, consider the following adversarial scenario: in a large-scale urban domain ($|\mathcal{R}| \rightarrow \infty$) where agents are interested only in resources that are physically close to them, the majority of resources would have utility $\approx 0$. If we select a resource in a weighted at random fashion, the probability of selecting a low utility resource would be high -- due to the large number of resources -- resulting in low social welfare.

In this work, we combine the aforedescribed two approaches. Let $\mathcal{N}^n$ denote the set of every possible agent that belongs to the same region of utility space as $n$, i.e., $\mathcal{N}^n = \{n' : u_{n'}(\cdot) \in \mathcal{D}_i \land u_n(\cdot) \in \mathcal{D}_j \Rightarrow i = j \}$. We refer to $\mathcal{N}^n$ as the set of neighbors of $n$. Note that the neighbors of an agent do not need to be in $\mathcal{N}$, we account for every potential agent (i.e., $\cup_{n \in \mathcal{N}} \mathcal{N}^n \supset \mathcal{N}$). The \emph{neighbors} are the \emph{set of agents that PLDP guarantees indistinguishability}. Then, each agent $n$ independently generates the sets ($\mathcal{R}^n_1, \dots, \mathcal{R}^n_i, \dots, \mathcal{R}^n_{R}$), where the set $\mathcal{R}^n_i$ contains the $i^{\text{th}}$ most preferred resource of each neighbor, i.e., $\mathcal{R}^n_i = \cup_{\forall n' \in \mathcal{N}^n} \{r_i^{n'}\}$.

\begin{table}[t!]
\centering
\caption{Nomenclature, Algorithm \ref{algo: palma}}
\label{tab:algorithm legend}
\resizebox{\columnwidth}{!}{%
\begin{tabular}{@{}ll@{}}
\toprule
$s$                                                   & Current step (indicates a specific set $\mathcal{R}^n_s$) \\
$g$                                                   & Specifies which resource to access          \\
\multirow{2}{*}{$\{Y, A_{r_1}, \dots, A_{r_{R}}\}$}   & $Y$ refers to yielding, and                 \\
                                                      & $A_r$ refers to accessing resource $r$      \\
$P_S^n(\cdot)$                                        & Resource selection probability distribution \\
$P_B^n(\cdot)$                                        & Back-off probability distribution           \\
$c$                                                   & Accumulated privacy cost                    \\
$c_{max}$                                             & Highest possible privacy cost for selection or back-off \\
$B_n$                                                 & Privacy budget                              \\ \bottomrule

\end{tabular}%
}
\end{table}

Agent $n$ moves in a \emph{sequential} manner from set to set (starting from the set of the most preferred resources, $\mathcal{R}^n_1$, and looping back to it after $\mathcal{R}^n_R$). The resource selection is performed in a \emph{weighted at random} fashion in the sets $\mathcal{R}^n_i$. Specifically, at step $s = t \mod R$, where $t$ is the current time-step, agent $n$ will select resource $r_i \in \mathcal{R}^n_s$ with probability given by (line \ref{line: selection prob} of Algorithm \ref{algo: palma}):
% \begin{equation} \label{Eq: selection probability}
%     P_S^n(i, s, \zeta_S) = (1 - \zeta_S) \frac{u_n(r_i)}{\sum_{r \in \mathcal{R}^n_s} u_n(r)} + \frac{\zeta_S}{|\mathcal{R}^n_s|}
% \end{equation}
\begin{equation} \label{Eq: selection probability}
    P_S^n(i, s, \zeta_S) = \zeta_S \times P_{\text{WaR}}(i, s, n) + (1 - \zeta_S) \times S_{\text{Noise}}(i, s, n^*)
\end{equation}
\begin{equation} \label{Eq: weighted at random}
    P_{\text{WaR}}(i, s, n) = \frac{u_n(r_i)}{\sum_{r \in \mathcal{R}^n_s} u_n(r)}
\end{equation}

Equation \ref{Eq: selection probability} defines a mixture distribution, composed of (a) selecting in a weighted at random fashion using the utilities of agent $n$ ($P_{\text{WaR}}(i, s, n)$, given by Equation \ref{Eq: weighted at random}), and (b) a distribution that introduces noise ($S_{\text{Noise}}(\cdot)$) to the selection process. $\zeta_S$ tunes the magnitude of the introduced randomness.

The introduced noise can be any distribution that is known and common for all agents (can be domain specific). For example, it could simply be a uniformly at random selection in the set of resources $\mathcal{R}^n_s$. In this work, we take advantage of domain knowledge. Specifically, let $n^*$ denote a `representative' agent of the Neighborhood of agent $n$. This can be for example a(n) (potential) agent located in the center of the neighborhood in a mobility-on-demand application. Then, the common distribution (i.e., noise) can be to play in a uniformly at random manner according to the utility function of the representative agent, i.e., $S_{\text{Noise}}(i, s, n^*) = P_{\text{WaR}}(i, s, n^*)$. In section \ref{sec: Elaborative Example on Neighborhoods}, we provide a concrete example on the fragmentation of the utility space into neighborhoods, the representative agent, and the selection and back-off probabilities.

\begin{algorithm}[!t]
  \caption{PALMA: Privacy-preserving ALtruistic MAtching.} \label{algo: palma}
  \begin{algorithmic}[1]
    \State \textbf{Initialize} $s \leftarrow 1$, $g \sim P_S^n(\cdot)$, $c \leftarrow 0$, $converged \leftarrow False$
    \State \textbf{Calculate} $c_{max}$ according to Equation \ref{Eq: max privacy cost}
        
    \Procedure{PALMA}{}
    \While{!converged}
      \If{ $g = A_r$}
        \State Agent $n$ attempts to acquire $r$
        \If{Collision($r$)}
          \If{$c + c_{max} \leq B_n$} \label{line: budget check 1}
            \State Back-off (set $g \leftarrow Y$) with probability $P_B^n(\cdot)$ \label{line: backoff prob}
            \State $c \leftarrow c + c_{max}$ \label{line: accounting 1}
          \Else
            \State Back-off (set $g \leftarrow Y$) with prob. $B_{\text{Noise}}(\cdot)$ \label{line: noisy backoff}
          \EndIf
        \Else
          \State converged $\leftarrow True$
        \EndIf
      \Else { ($g = Y$)}
          \State $s \leftarrow (s + 1) \mod R$
          \If{$c + c_{max} \leq B_n$} \label{line: budget check 2}
            \State Agent $n$ monitors $r \sim P_S^n(\cdot)$  \label{line: selection prob}
            \State $c \leftarrow c + c_{max}$ \label{line: accounting 2}
          \Else
            \State Agent $n$ monitors $r \sim S_{\text{Noise}}(\cdot)$ \label{line: noisy selection}
          \EndIf
          \If{Free($r$)} set $g \leftarrow A_r$
          \EndIf
      \EndIf
    \EndWhile
    \EndProcedure

    \State \textbf{Output} $r$, such that $g = A_r$, and $(\varepsilon, \delta) \leftarrow$ getPrivacy($c$) (Eq.\ref{Eq: get privacy})
  \end{algorithmic}
\end{algorithm}

\subsubsection{\textbf{Back-off Distribution}} \label{sec: Back-off Distribution}

The back-off probability, $P_B^n(\cdot)$ (line \ref{line: backoff prob} of Algorithm \ref{algo: palma}), is computed individually and locally based on each agent's expected utility loss that he will incur if he switches:
\begin{equation} \label{Eq: loss}
  loss(i, s, n) = u_n(r_i) - \underset{r_j \in \mathcal{R}^n_{s + 1}}{\sum} \frac{u_n(r_j)}{\sum_{r \in \mathcal{R}^n_{s + 1}} u_n(r)} u_n(r_j)
\end{equation}

\noindent
The actual back-off probability can be computed with any monotonically decreasing function $f$ on $loss(\cdot)$, e.g.:
\begin{equation} \label{Eq: linear}
  f(loss) =
  \begin{cases}
    1 - \gamma, & \text{ if } loss \leq \gamma \\
    \gamma, & \text{ if } 1 - loss \leq \gamma \\
    1 - loss, & \text{ otherwise}
  \end{cases}
\end{equation}

\noindent
where $\gamma$ places a threshold on the minimum / maximum back-off probability. According to the above distribution, agents that do not have good alternatives will be less likely to back-off and vice versa. The ones that do back-off select an alternative resource, according to the resource selection probability $P_S^n(\cdot)$, and examine its availability (line \ref{line: selection prob} of Algorithm \ref{algo: palma}). Finally, $P_B^n(\cdot)$ is given by Equation \ref{Eq: backoff probability}:
\begin{equation} \label{Eq: backoff probability}
  P_B^n(i, s, \zeta_B) = \zeta_B \times f(loss(i, s, n)) + (1 - \zeta_B) \times B_{\text{Noise}}(i, s, n^*)
\end{equation}

The back-off distribution is mixture between acting according to an agent's own utility function ($f(loss(i, s, n))$), and a distribution that introduces noise ($B_{\text{Noise}}(\cdot)$) to the back-off process. $\zeta_B$ tunes the magnitude of the introduced randomness. As was the case with the selection distribution, the introduced noise for the back-off distribution can be any distribution that is known and common for all agents. In this work, we set $B_{\text{Noise}}(i, s, n^*) = f(loss(i, s, n^*))$, i.e., the `noise' distribution refers to backing-off according to the utility function of the `representative' agent (described in Section \ref{sec: Resource Selection Distribution}).

% trim={<left> <lower> <right> <upper>}
\begin{figure}[t!]
    \centering
    \includegraphics[width = 1 \linewidth, height = 10em, trim={4em 5em 4em 8em}, clip]{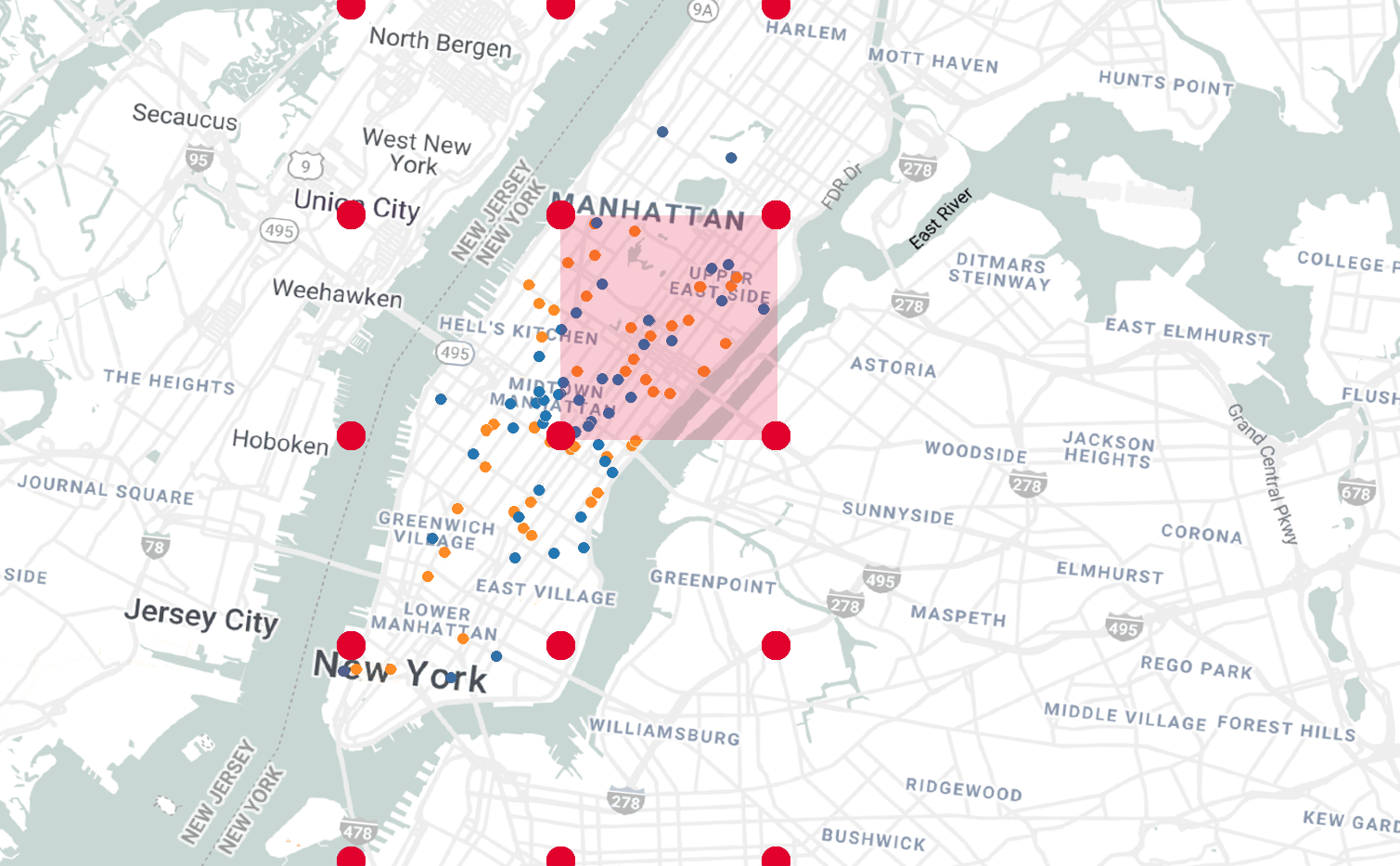}
    \caption{A visual representation of the regions ($\{ \mathcal{D}_i \}$) of PLDP for the mobility-on-demand application. Red dots denote the edge points of each region ($\ell = 4000$). Orange dots represent the agents (requests), and blue dots represent the resources (vehicles) in our dataset. As an example, an agent in the overlaid rectangle could be located \emph{anywhere} in the rectangle from the attacker's point of view.}
    \Description{A visual representation of the regions of Piecewise Local Differential Privacy.}
    \label{fig: map}
\end{figure}

\subsubsection{\textbf{Elaborative Example on Neighborhoods}} \label{sec: Elaborative Example on Neighborhoods}

In what follows, along with Section \ref{sec: Elaborative Example on Privacy cost}, we will provide an elaborative, practical example of the key notions of PALMA.

PLDP is used to protect the utility function of agents. Consider the space of all possible utility functions, and then consider the space of the images of those utility functions. We fragment the former into sub-spaces $D_i$, such that for two utility functions that belong to the same $D_i$, their image is `close' in distance. In simple terms this means that the actual utility value of a resource would be similar for agents with utility functions in the same sub-space $D_i$. The fragmentation is performed by $\phi(\cdot)$.

Each agents selects his own $\phi(\cdot)$ based on his privacy needs. The choice of $\phi(\cdot)$ is public information. For simplicity, in this work, we assume that every agent has the same $\phi(\cdot)$. The choice of $\phi(\cdot)$ fragments the space of agents into regions; the image of the utility function of every agent in a region is close in distance to every other agent in the same region. The definition of the region ($\phi(\cdot)$) as well as the distance metric are domain specific.

As a concrete example, consider a mobility-on-demand (MoD) application (e.g., ridesharing). Let the utility of each agent (ridesharing user) be inversely proportional to the distance (in meters) from the resource (vehicle). In this case, we can split the are of operation into rectangular regions, as shown in Figure \ref{fig: map}; agents in the same region would have similar utilities for each resource.\footnote{Note that we protect the privacy of the agents, not the resources; thus, the resources (vehicles) do not need to belong to any region, and can be matched with any agent regardless of his region.}

% \footnote{The utility function may use multiple inputs (e.g., distance, price, etc.) to calculate the utility value of a resource.}
% \footnote{For the paper assignment problem, following the related literature, we use the cosine similarity as a distance metric (see Supplementary Material).}

To compute his neighbors, an agent considers \emph{every possible agent} that could belong in his region, regardless if this agent exists. Expanding on our MoD example, we can consider having an agent ridesharing user) every, e.g., 10m on the map. In a $10^6$m$^2$ region, the neighborhood will include $10^4$ agents. Each of these agents has his own preference (ordering) of resources. Using these preferences, we can construct the sets $\mathcal{R}_1^n, \dots, \mathcal{R}_R^n$, where the set $\mathcal{R}^n_i$ contains the $i^{\text{th}}$ most preferred resource of each neighbor. The construction of the neighborhoods needs to be performed once, offline. PLDP guarantees that each agent is \emph{indistinguishable from all his neighbors} (i.e., every potential agent that could exist in his region) from the attacker's point of view.  

Finally, the `representative' agent of each region can be a `virtual' agent located at the center of the region. Given that $\phi(\cdot)$ is public -- and thus the fragmentation into regions as well -- the selection and back-off distribution of the representative agent is also public and common for all agents.

% \subsection{Complexity} \label{Complexity}

% \subsubsection{\textbf{Bounding the Set of Desirable Resources}} \label{sec: Resource Set}

\subsection{\textbf{Communication and Computation Complexity}} \label{sec: Communication and Computation Complexity}

PALMA (just like ALMA~\cite{ijcai201931}) does not require any inter-agent communication\footnoteref{footnote: no communication}. The initialization is linear to the size of the region, $\mathcal{O}(\max_i|\mathcal{D}_i|)$, but this can be done once off-line. The accounting of the privacy loss is $\mathcal{O}(1)$. Finally, PALMA converges in polynomial time in the general case, and in \emph{constant} time in the realistic case where each agent is interested in a subset of the total resources (i.e., $\mathcal{Q}^n \subset \mathcal{R}$) and thus at each resource there is a bounded number of competing agents ($\mathcal{V}^r \subset \mathcal{N}$) (see Appendix \ref{supp: PALMA supp}).

\subsection{Privacy Mechanism} \label{sec: Privacy Mechanisms}

% PALMA implements two separate defense mechanisms, applied in three different parts of the algorithm, all of which can be separately tuned depending on the desired level of privacy.% at each stage.

PALMA's defense mechanism is based on the idea of randomized response~\cite{warner1965randomized}, and involves adding controlled randomness in (i) the resource selection and (ii) back-offs, parametrized by $\zeta_S$ and $\zeta_B$, respectively (see Equation \ref{Eq: selection probability} and \ref{Eq: backoff probability}). The idea is that the agent first flips a coin to decide whether to act truthfully. Then, with probability $\zeta_S$ (or $\zeta_B$), the agent plays according to its true selection (or back-off) function; with probability $1-\zeta_S$ (or $1-\zeta_B$), the agent plays according to a public, common distribution.

Moreover, each agent has a privacy budget of $\varepsilon = B_n$. Upon depletion in the course of using the above mechanisms (see lines \ref{line: budget check 1} \& \ref{line: budget check 2} of Algorithm \ref{algo: palma}), the agent will play \emph{noisy} actions (see lines \ref{line: noisy backoff} \& \ref{line: noisy selection} of Algorithm \ref{algo: palma}). Note also that each agent can select the fragmentation function $\varphi(\cdot)$ of PLDP and adjust the size of the neighborhood $\mathcal{N}^n$ according to his privacy needs.

\section{Privacy Accounting}
\label{sec:pldp_accounting}
Since PALMA is an iterative algorithm, we need to compute $(\varepsilon, \delta)$ guarantees over multiple applications of the privacy mechanism. This can be done via \emph{privacy accounting} methods (e.g., \cite{dwork2014algorithmic}). We employ the accounting framework introduced in \cite{triastcyn2020bayesian} and extend it to generic subsampled mechanisms. While developed for the notion of Bayesian DP, this framework is applicable to the traditional DP as well, and in such a case, is equivalent to the moments accountant~\cite{abadi2016deep} for the subsampled Gaussian mechanism and R\'enyi accountant~\cite{mironov2017renyi}. Let us briefly outline the method.

% The most simple of these methods is to use the basic composition theorem of DP~\cite{dwork2014algorithmic}. More advanced tools have been developed in the machine learning community~\cite{abadi2016deep,mironov2017renyi}, but they have largely focused on the Gaussian noise mechanism in particular.

Let $\sigma_t$ and $\sigma'_t$ denote signals sent by agents $x$ and $x'$ in time-step $t$, and $\xi_t$ any auxiliary information. A set of signals (auxiliary information) sent in time-steps $1$ through $T$ is denoted by $\sigma_{1:T}$ ($\xi_{1:T}$). In the context of PALMA, these signals represent either an attempt to acquire a resource, or a back-off from a previously contested resource\footnote{In an arbitrary domain, the signal would correspond to an action of an agent.}, while the auxiliary information corresponds to $s$ (which determines the set of resources $\mathcal{R}_s$, see Equation \ref{Eq: selection probability}, \ref{Eq: backoff probability}).  Following \cite{triastcyn2020bayesian}, we also introduce the notion of \emph{privacy cost}:
\begin{equation*}
c_t(\sigma_t, \xi_t, x, x', \lambda) \triangleq \max
\begin{cases}
\lambda \mathcal{D}_{\lambda+1} [p(\sigma_t | \xi_t, x) \| p(\sigma_t | \xi_t, x')] \\
\lambda \mathcal{D}_{\lambda+1} [p(\sigma_t | \xi_t, x') \| p(\sigma_t | \xi_t, x)]
\end{cases}
\end{equation*}

\noindent
where $\mathcal{D}_\lambda (\cdot \| \cdot)$ is the R\'enyi divergence of order $\lambda$ (see App.~\ref{supp: Renyi Divergence Definition} ).
% Section \ref{supp: Renyi Divergence Definition}.

% \begin{equation*}
% \label{eq:privacy_cost}
%   c_t(\sigma_t, \xi_t, x, x', \lambda) \triangleq \lambda \mathcal{D}_{\lambda+1} [p(\sigma_t | \xi_t, x) \| p(\sigma_t | \xi_t, x')].
% \end{equation*}
% Finally, we can formulate the theorem for accounting general subsampled mechanisms.

% \begin{theorem}
% \label{thm:general_subsampled_privacy_cost}
% Given the subsampling probability $q$, the privacy cost of one application of privacy mechanism (a single round $t$ of PALMA) for $\lambda \in \mathbb{N}$ can be computed as
% \begin{equation*}
%   c_t(\sigma_t, \xi_t, x, x', \lambda) = \max\{c_t^{L}(\sigma_t, \xi_t, x, x', \lambda), c_t^{R}(\sigma_t, \xi_t, x, x', \lambda)\},
% \end{equation*}
% where
% \begin{align*}
%   &c_t^{L}(\sigma_t, \xi_t, x, x', \lambda) = \log \mathbb{E}_{k \sim B(\lambda+1, q)} \left[ \mathbb{E}_{s_t} \left[ \left(\frac{p(\sigma_t | \xi_t, x')}{p(\sigma_t | \xi_t, x)}\right)^k \right] \right], \\
%   &c_t^{R}(\sigma_t, \xi_t, x, x', \lambda) = \log \mathbb{E}_{k \sim B(\lambda+1, q)} \left[ \mathbb{E}_{s_t} \left[ \left(\frac{p(\sigma_t | \xi_t, x)}{p(\sigma_t | \xi_t, x')}\right)^k \right] \right],
% \end{align*}
% and $B(\lambda, q)$ is the binomial distribution with $\lambda$ experiments and probability of success $q$.
% \end{theorem}
% \begin{proof}
% See the supplementary material.
% \end{proof}

\subsection{PALMA's Privacy Cost} \label{sec: PALMA's Privacy Cost}

Every matching game starts with a fresh set of agents with random identifiers. Each agent computes (\emph{once}, and \emph{off-line}) the highest possible privacy cost at any round ($c_{max}$), i.e., the maximum value between the worst possible privacy cost during resource selection and back-off:
\begin{equation} \label{Eq: max privacy cost}
c_{max} = \max
\begin{cases}
\underset{\xi_t \in \{1, \dots, R\}}{\max}  \; \underset{x' \in \mathcal{N}^x}{\max} \; \underset{\sigma_t \in \mathcal{R}^x_{\xi_t} \sim P_S^n(\cdot)}{\max} \; c_t(\cdot) \\
\underset{\xi_t \in \{1, \dots, R\}}{\max}  \; \underset{x' \in \mathcal{N}^x}{\max} \; \underset{\sigma_t \in \mathcal{R}^x_{\xi_t} \sim P_B^n(\cdot)}{\max} \; c_t(\cdot)
\end{cases}
\end{equation}

The agents do not change their utilities during the matching process (i.e., the distributions $P_S^n(\cdot)$ and $P_B^n(\cdot)$ stay fixed), thus each agent can \emph{compute a priori} the total privacy cost (\emph{worst case privacy guarantees}) and the maximum number of rounds until the budget $B_n$ is exhausted and he has to play according to the noise distributions. Agents can then adjust their privacy parameters accordingly. The actual privacy loss is accounted on the fly during execution (see lines \ref{line: accounting 1} and \ref{line: accounting 2} of Algorithm \ref{algo: palma}).

To bound the total privacy loss over multiple rounds and compute $\varepsilon$ from $\delta$ or vice versa, we can use an advanced composition theorem. As stated, the advanced compositions theorem for the Bayesian accountant~\cite{triastcyn2020bayesian}, the moments accountant~\cite{abadi2016deep} and the R\'enyi accountant~\cite{mironov2017renyi} are equivalent in this case, resulting in:
\begin{minipage}{.5\linewidth}
\centering
\begin{equation*}
  \log \delta \leq \sum_{t=1}^T c_{max}(\cdot) - \lambda \varepsilon
\end{equation*}
\end{minipage}%
\begin{minipage}{.5\linewidth}
\centering
\begin{equation*} \label{Eq: get privacy}
  \varepsilon \leq \frac{1}{\lambda} \sum_{t=1}^T c_{max}(\cdot) - \frac{1}{\lambda} \log \delta
\end{equation*}
\end{minipage}

% \begin{align}
% \label{eq:delta_from_eps}
%   \log \delta &\leq \sum_{t=1}^T c_t(\lambda) - \lambda \varepsilon, \\
% \label{eq:eps_from_delta}
%   \varepsilon &\leq \frac{1}{\lambda} \sum_{t=1}^T c_t(\lambda) - \frac{1}{\lambda} \log \delta.
% \end{align}

It is important to note that the above $\varepsilon$ and $\delta$ should not be published, since the agent uses his own utility function to calculate the cost (in Equation \ref{Eq: max privacy cost}).

\subsubsection{\textbf{Elaborative Example on the Privacy Cost Calculation}} \label{sec: Elaborative Example on Privacy cost}

In this section we expand on our practical example on MoD systems introduced in Section \ref{sec: Elaborative Example on Neighborhoods}. 

Recall that PLDP provides \emph{Local} DP guarantee, meaning a bound on the outcome probabilities for \emph{any} pair of individual agents, inside the region. As such, to compute the privacy cost per round, each agent $n$ has to identify the neighbors that would result to the maximum privacy loss (i.e., their selection (back-off) distributions result in the largest R\'enyi divergence, see Equation \ref{Eq: max privacy cost}). Thus, each agent $n$ independently identifies two agents $n'$, and $n''$ \emph{from his neighborhood} that result in the worst privacy loss given the agent's selection and back-off distributions (Equation \ref{Eq: selection probability} and \ref{Eq: backoff probability}, respectively). Then, he can compute the \emph{worst case} privacy loss in any round by taking the maximum of the two values (Equation \ref{Eq: max privacy cost}). Using this information, each agent is able to (i) compute his total privacy cost \emph{a priori} and adjust his privacy parameters accordingly, (ii) keep track of his privacy budget at every time-step, and (iii) calculate his total $\varepsilon$ after convergence. This process needs to happen \emph{once}, \emph{offline}. As mentioned, each agent can adjust the size of the neighborhood $\mathcal{N}^n$ (e.g., length $\ell$, see Section \ref{sec: Mobility on Demand - Setting}) according to his privacy needs.

\section{Evaluation} \label{sec: Evaluation}

We evaluate PALMA in a \emph{mobility-on-demand} and a \emph{paper assignment} application, using \emph{real-data} for both. We focus on the social welfare (sum of utilities, $\sum_{n \in \mathcal{N}} u_n(\cdot)$) and level of privacy ($\varepsilon$ given $\delta = 10^{-5}$). Each problem instance is run $32$ times. We report the average value for the social welfare, the average value for the median of $\varepsilon$, and the maximum value of $\varepsilon$. Error bars represent one standard deviation. We set $\zeta_S = 0.2$, $\zeta_B = \gamma = 0.05$, $B_n = 1$, $\lambda = 32$.
% \footnote{For the MoD application. See the supplement for the paper assignment.}

\section{Test-Case 1: Mobility-on-Demand} \label{sec: Mobility on Demand}

\subsection{Motivation} \label{sec: Mobility on Demand - Motivation}

The emergence and widespread use of mobility-on-demand (MoD) services (e.g., ridesharing platforms like Uber or Lyft) in recent years has had a profound impact on urban transportation. Normally the process is facilitated by a centralized operator, that requires accurate location information of passengers and vehicles, which raises privacy concerns. Such a problem is ideal to showcase PALMA, as explained in Section \ref{PLDP Motivation}. Moreover, contrary to other approaches (e.g.,~\cite{prorok2017privacy,10.5555/3237383.3237910}) PALMA is \emph{decentralized} and employs \emph{Local DP}, providing privacy against a malicious data curator.

The Ridesharing and Fleet Relocation problem can be decomposed into three weighted matching sub-problems, all of which can be solved efficiently by ALMA~\cite{danassis2019putting} (and thus by PALMA as well). In this test-case we will focus on passenger to vehicle matching, using PLDP and PALMA to provide a \emph{scalable}, \emph{on-device}, decentralized solution that \emph{protects user preferences} (user location in this context).

\subsection{Setting} \label{sec: Mobility on Demand - Setting}

% trim={<left> <lower> <right> <upper>}
\begin{figure}[t!]
    \centering
    \includegraphics[width = 1 \linewidth, trim={0em 1.1em 0em 0.8em}, clip]{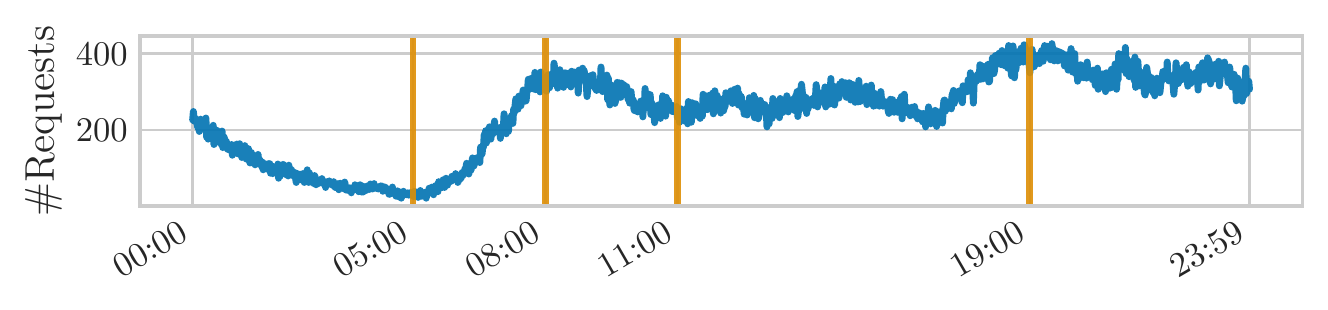}
    \caption{Request per minute in Manhattan on Jan. 15, 2016. Vertical lines denote the selected evaluation instances.}
    \Description{Request per minute in Manhattan on January 15, 2016.}
    \label{fig: requests per minute}
\end{figure}

Our evaluation setting is specifically designed to \emph{resemble reality as closely as possible}, following the modeling of~\cite{danassis2019putting}. We have used the NYC yellow taxi trip records~\cite{TLCTripRecordData}. For every request, the dataset provides amongst others the geo-location coordinates.

We report results on four 30s instances on a typical day (Jan 15th). These instances were selected to represent various distributions of demand (see Figure \ref{fig: requests per minute}): the two highest peaks, the lowest peak, and a mid-day low\footnote{Specifically, 05:00:00 - 05:00:30 represents the lowest demand, 08:00:00 - 08:00:30 and 19:00:00 - 19:00:30 represent the two rush hours (in the morning and evening, respectively), and finally, 11:00:00 - 11:00:30 represents a mid-day low.}. We selected 30s periods because in practice the granularity of in-batches approaches for MoD services is between\footnote{We also ran the same instances in batches of 10s and obtained \emph{better results} (in terms of social welfare), but opted to present the worst case.} 10s to 30s~\cite{alonso2017demand,ijcai2020-609,prorok2017privacy,danassis2019putting}. It is important to stress this \emph{does not affect the scalability} of the proposed approach. Running PALMA for a day, for example, would simply result in running $24 \times 60 \times 2$ batches (as was done in~\cite{danassis2019putting}). Assuming similar distributions for requests and vehicles\footnote{A reasonable assumption given that our choice of evaluated distributions covers all the extremes, and a typical mid-day demand.}, the social welfare and privacy cost of each agent will remain approximately the \emph{same}, since the privacy cost (Equation \ref{Eq: max privacy cost}) \emph{only depends on the size of the region $\mathcal{D}_i$}.

The set of agents $\mathcal{N}$ is composed by the requests in Manhattan ($17$, $154$, $116$, and $174$ requests in total on each of the evaluated batches). The set of resources $\mathcal{R}$ includes an equal number of vehicles scattered across the map. To avoid cold start, the position of each of the vehicles was set to the drop-off geo-location of the last (prior to the start time of the simulation) $x$ requests (where $x$ is the number of vehicles in each case). We used the Manhattan distance as a distance function (using the Haversine formula\footnote{\url{https://en.wikipedia.org/wiki/Haversine_formula}} to calculate the distance in each coordinate), as it has been found to be a close approximation of the actual driving distance in Manhattan~\cite{danassis2019putting}. The utility function is $u_n(r) = e^{-\frac{d(n, r)}{\alpha}}$, where $\alpha = 4000$ controls the steepness and $d(n, r)$ denotes the distance between agent $n$ and resource $r$ (in m). We opted to use an exponential function to enable short pick-up times, as research conducted by ridesharing companies shows that a short pick-up time is important for passengers' satisfaction~\cite{8259801,lyftblog2}.

% where $\alpha$ controls the steepness in utility loss as we increase the distance between an agent and a resource and $d(n, r)$ denotes the distance between agent $n$ and resource $r$.

% \begin{equation} \label{eq: taxis utility}
%   u_n(r) = e^{-\frac{d(n, r)}{\alpha}}
% \end{equation}

The map is divided into \emph{fixed} square regions of edge length $\ell$ (which correspond to the $\mathcal{D}_i$). PLDP demands that a user is indistinguishable, from the attacker's point of view, from any potential user that could exist in the same region\footnote{We assume that potential neighbors are 100m apart in every direction.} (i.e., all his neighbors, see Sections \ref{sec: Resource Selection Distribution} and \ref{sec: Elaborative Example on Neighborhoods}). We have evaluated $\ell \in \{1000, 2000, 3000, 4000\}$ m, which roughly correspond to an area of $\{45.6, 182.5, 410.5, 730\}$ city blocks\footnote{The standard city block in Manhattan is about 80 m $\times$ 274 m (\url{https://en.wikipedia.org/wiki/City_block}).}. Figure \ref{fig: map} offers a visual representation of the setting.

\subsection{Baselines} \label{sec: Baselines}

We employ the centralized Hungarian algorithm~\cite{kuhn1955hungarian} to compute the non-private maximum-weight -- i.e., optimal in terms of social welfare -- solution, which we use to compare the loss in social welfare of all of the evaluated algorithms. We compare PALMA against three privacy-preserving baselines:
\begin{enumerate}
  \item The Hungarian algorithm~\cite{kuhn1955hungarian} -- which is an optimal assignment centralized algorithm -- made private by obfuscating (adding noise) the geo-location coordinates according to geo-indistinguishability~\cite{andres2013geo} (similarly to~\cite{prorok2017privacy}).
  \item The original ALMA~\cite{ijcai201931} under similarly obfuscated (noisy) geo-location coordinates\footnote{Note that we also attempted to use the original ALMA with Local Differential Privacy, yet, due to the large problem size, the privacy budget only sufficed for one round.}.
  \item The maximally private solution (i.e., the centralized random).
\end{enumerate}

For the geo-indistinguishability-based baselines, we calculated a noisy geo-location for each agent and resource, according to Algorithm \ref{tb: geo-ind algo}, which can be found in the appendix.
% \ref{tb: geo-ind algo}

% \subsection{Simulation Results} \label{sec: Mobility on Demand - Simulation Results}

% We compare PALMA against three baselines: (i) the Hungarian algorithm~\cite{kuhn1955hungarian} -- which is an optimal assignment centralized algorithm -- made private by obfuscating (adding noise) the geo-location coordinates according to geo-indistinguishability~\cite{andres2013geo} (similarly to~\cite{prorok2017privacy}), (ii) the original ALMA~\cite{ijcai201931} under similarly obfuscated (noisy) geo-location coordinates\footnote{Note that we also attempted to use the original ALMA with Local Differential Privacy, yet, due to the large problem size, the privacy budget only sufficed for one round (after which the agents had to play randomly).}, and (iii) the maximally private solution (i.e., the centralized random). Please see the supplement for more information.

\subsection{Simulation Results: Social Welfare} \label{sec: Mobility on Demand - Social Welfare}

\begin{figure}[t!]
    \centering
    \includegraphics[width = 1 \linewidth, trim={0.8em 0.8em 0.8em 0.9em}, clip]{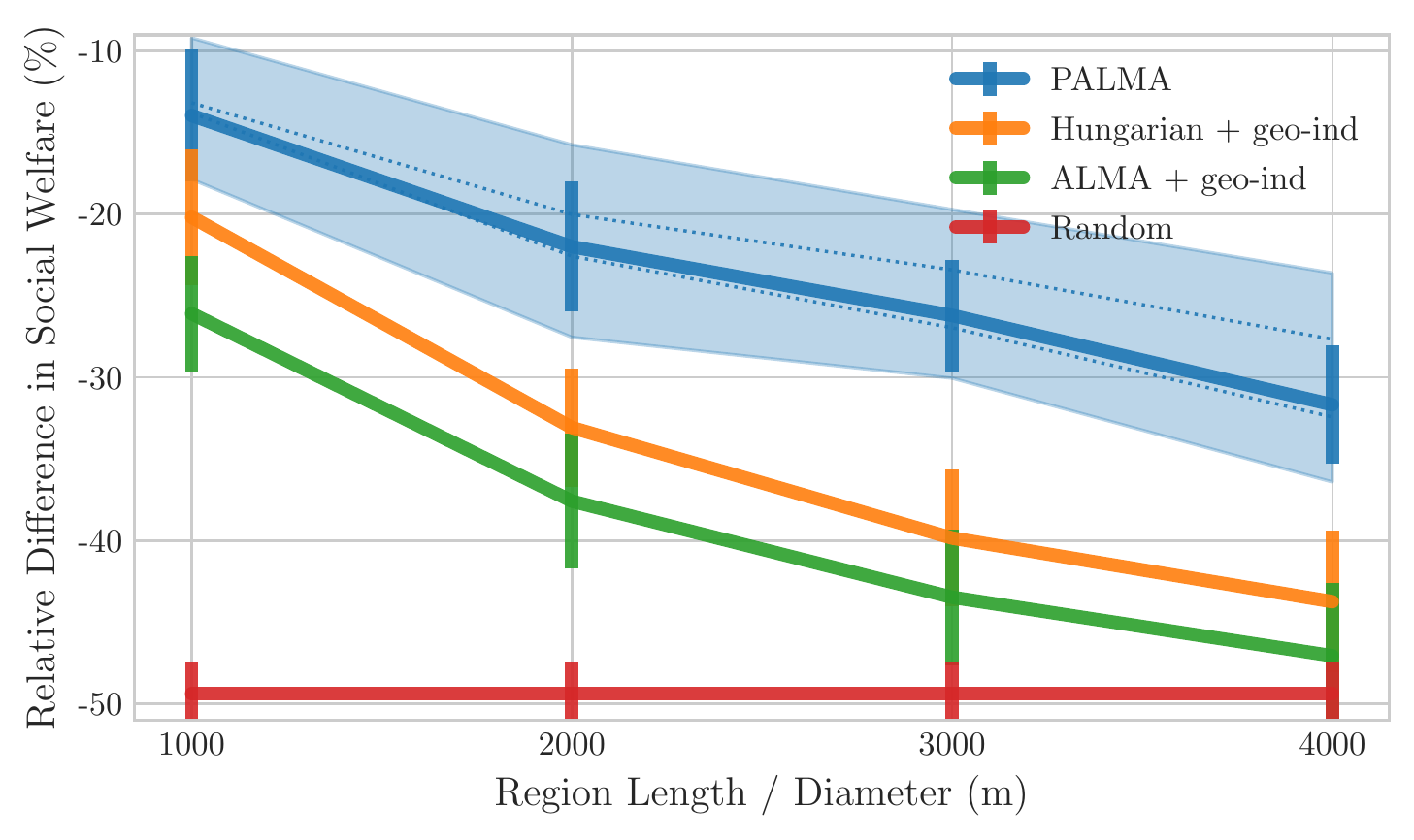}
    \caption{Loss in SW compared to the non-private, optimal solution for increasing region edge length ($\ell$) and $\varepsilon = 1$. The dotted lines represent the upper ($\varepsilon \rightarrow \infty$) and lower ($\varepsilon = 0$) bound for PALMA, while the shaded area adds one standard deviation to the aforementioned bounds (see Section \ref{sec: Mobility on Demand - Social Welfare}).}
    \Description{Loss in social welfare compared to the non-private, optimal solution.}
    \label{fig: socialWelfare}
\end{figure}

For $\varepsilon = B_n = 1$ given $\delta = 10^{-5}$ (Figure \ref{fig: socialWelfare}), PALMA loses between $13.9 \pm 4.1\%$ ($\ell = 1000$) to $31.7 \pm 3.6\%$ ($\ell = 4000$) in social welfare compared to the non-private, optimal solution. The dotted lines represent the upper and lower bound; the upper bound assumes infinite budget ($B_n \rightarrow \infty$) thus the agents play according to their own utilities ($\zeta_S = \zeta_B = 1$), while the lower bound assumes zero budget ($B_n = 0$) thus the agents play according to the noise distribution ($\zeta_S = \zeta_B = 0$), i.e., according to the utilities of the representative agent. The shaded area adds one standard deviation to the aforementioned bounds.

For the same $\varepsilon$ guarantee and the same length as the privacy diameter, Hungarian + geo-ind loses between $20.2 \pm 4.2\%$ to $43.7 \pm 4.3\%$, while ALMA + geo-ind loses between $26.1 \pm 3.5\%$ to $47.1 \pm 4.5\%$. Finally, the maximally private solution (i.e., the centralized random), losses $49.4 \pm 2\%$.

% palma_eps_1_sw_loss_mean =              [-13.9 -22.  -26.2 -31.7]
% palma_eps_1_sw_loss_sd =                [4.1 4.  3.4 3.6]

% geoInd_Hungarian_eps_1_sw_loss_mean =   [-20.2 -33.1 -39.8 -43.7]
% geoInd_Hungarian_eps_1_sw_loss_sd =     [4.2 3.6 4.2 4.3]

% alma_geoInd_eps_1_sw_loss_mean =        [-26.1 -37.6 -43.5 -47.1]
% alma_geoInd_eps_1_sw_loss_sd =          [3.5 4.1 4.2 4.5]

PLDP and the carefully crafted noise of PALMA, allows PALMA to \emph{outperform even the centralized optimal} solution (Hungarian + geo-ind) by $27.6\%$ ($\ell = 4000$) to $30.9\%$ ($\ell = 1000$). In fact, if we increase the privacy requirement to $\varepsilon = 0.75$, the improvement increases to $31.3\%$ ($\ell = 4000$) to $45.9\%$ ($\ell = 1000$). Note that, besides the higher social welfare for the same privacy guarantee, PALMA is inherently decentralized and orders of magnitude faster than the Hungarian.

% \note{Note that the solution quality will degrade for LDP, as the problem size grows.}

% In this test-case, every agent is interested in all the resource, i.e., $\mathcal{Q}^n = \mathcal{R}, \forall n \in \mathcal{N}$, but in a real-world scenario, there would be a bound to the size of the set of resources each agent is interested in (see supplement). The benefit would be twofold: (i) PALMA will converge in \emph{constant} time, and (ii) if agents are only interested in resources in their vicinity, the social welfare will remain approximately \emph{constant}, even if we run PALMA in \emph{unboundedly large settings} (e.g., an entire country), assuming a fixed privacy region length ($\ell$).

% trim={<left> <lower> <right> <upper>}

\subsection{Simulation Results: Privacy} \label{sec: Mobility on Demand - Privacy Loss}

% While all the evaluated approaches offer the same ex-ante worst case privacy guarantee, PALMA offers a stronger ex post result. In PALMA, every agent has a budget $\varepsilon = B_n$ and can compute a priori the maximum number of rounds until the budget is exhausted and he has to play randomly (see Section \ref{sec: PALMA's Privacy Cost}). Yet, during runtime most agents converge at a smaller number of rounds, resulting in a small number of privacy mechanism invocations, and thus offering a stronger privacy guarantee to most agents (compared to the ex-ante $\varepsilon$, which is what, e.g., Hungarian or ALMA + geo-ind offer).

While the worst-case guarantee is the same across the evaluated methods, PALMA yields a stronger result on a per-agent basis. In PALMA, every agent has a budget $\varepsilon = B_n$ and can compute a priori the maximum number of rounds until the budget is exhausted and he has to play according to the noise distributions (see Section \ref{sec: PALMA's Privacy Cost}). During runtime, though, most agents converge in a few rounds (i.e, few privacy mechanism invocations), thus  accumulating smaller privacy loss compared to geo-ind based methods.

\begin{figure}[t!]
    \centering
    \includegraphics[width = 1 \linewidth, trim={0.8em 0.8em 0.5em 0.8em}, clip]{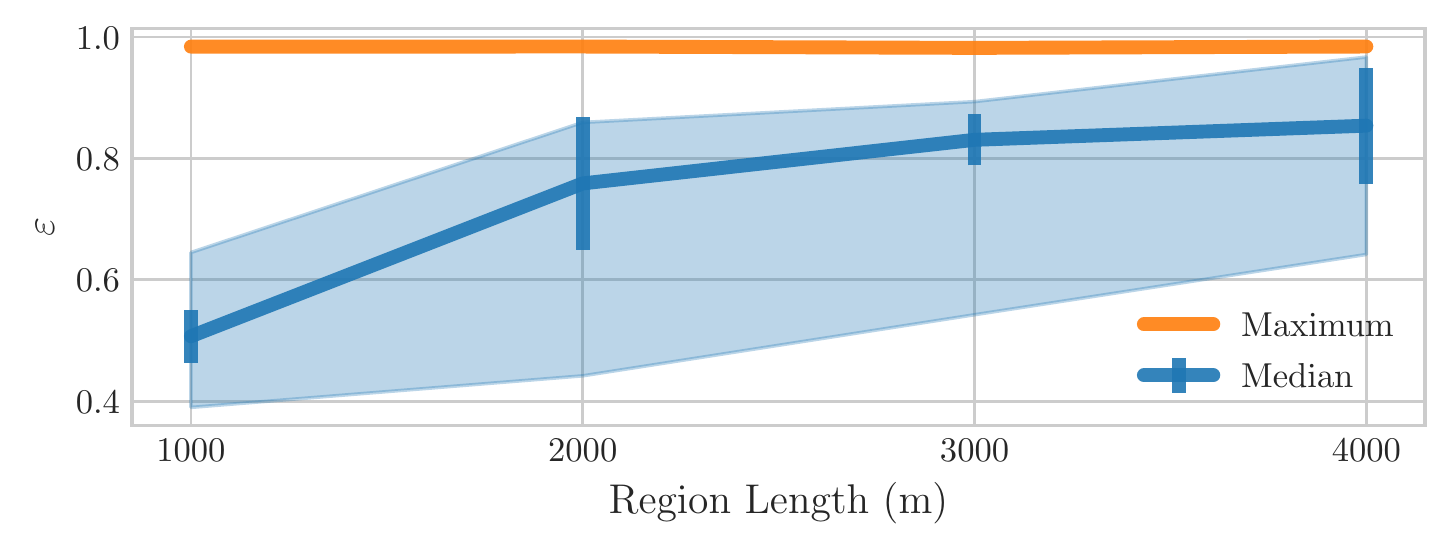}
    \caption{Maximum (orange) and median (blue) per-agent $\varepsilon$ for increasing region length ($\ell$). The shaded area represents the range between the max and min value of the median.}
    \Description{Maximum and median privacy loss.}
    \label{fig: privacyLoss}
\end{figure}

To demonstrate the latter, Figure \ref{fig: privacyLoss} depicts the maximum (out of all the $32$ runs) and median (average median value over the $32$ runs) per-agent $\varepsilon$ for increasing values of privacy region length $\ell$. PALMA is able to achieve a \emph{strong level of privacy} even in large-scale simulations. The average value of the median for $\ell = 1000$ is only $0.5$. Of course, the maximum per-agent $\varepsilon$ is bounded by the privacy budget (i.e., $\varepsilon = 1$). Recall that $\ell = 1000$ m corresponds to an area of $45.6$ city blocks, and $\ell = 4000$ m is larger than the width of Manhattan (which is 3700 m wide at its widest).
% ; intuitively this number means that an \emph{all-powerful attacker} can `distinguish' an individual with probability of only~\cite{triastcyn2020data} $\frac{1}{1 + e^{-\varepsilon}} = 0.62$, meaning practically random

\begin{figure}[t!]
  \centering
  \begin{subfigure}[t]{0.24\textwidth}
    \centering
    \includegraphics[width = 1 \linewidth, trim={0.7em 0.7em 0.75em 0.7em}, clip]{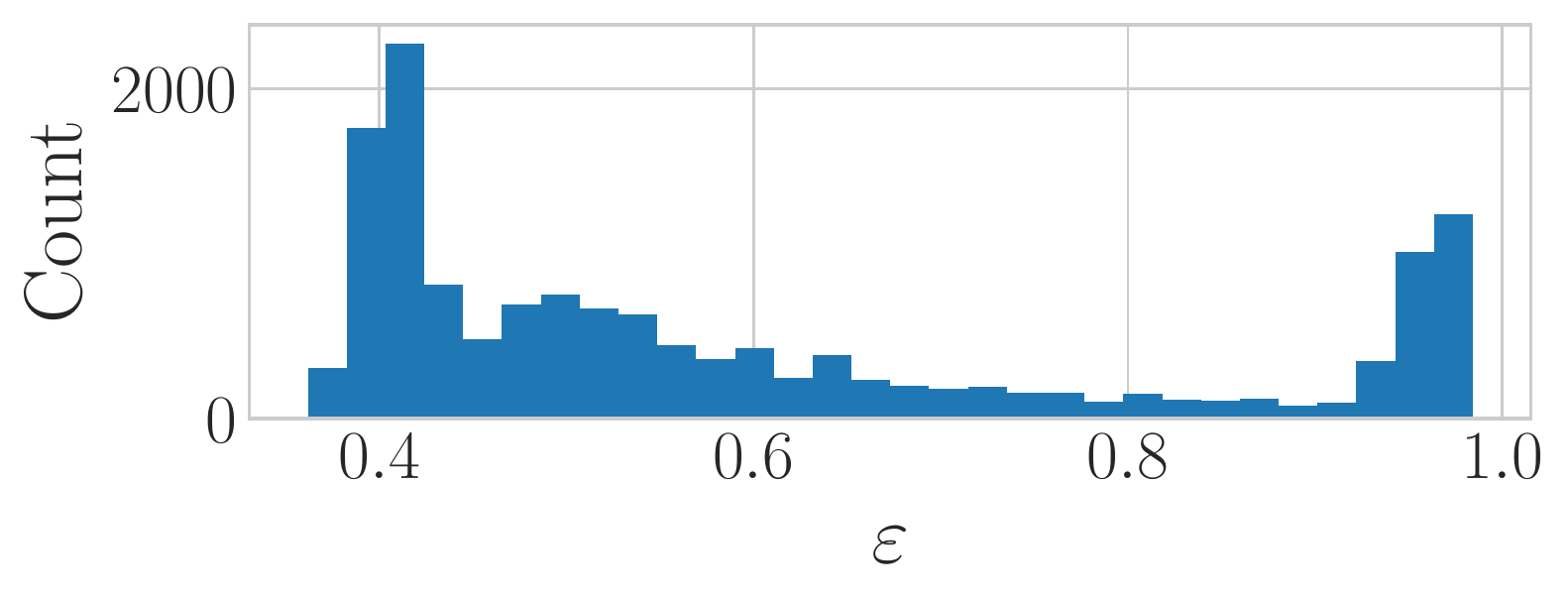}
    \caption{$\ell = 1000$}
    \label{fig: histogram_500}
  \end{subfigure}%
  ~ 
  \begin{subfigure}[t]{0.24\textwidth}
    \centering
    \includegraphics[width = 1 \linewidth, trim={0.7em 0.7em 0.75em 0.7em}, clip]{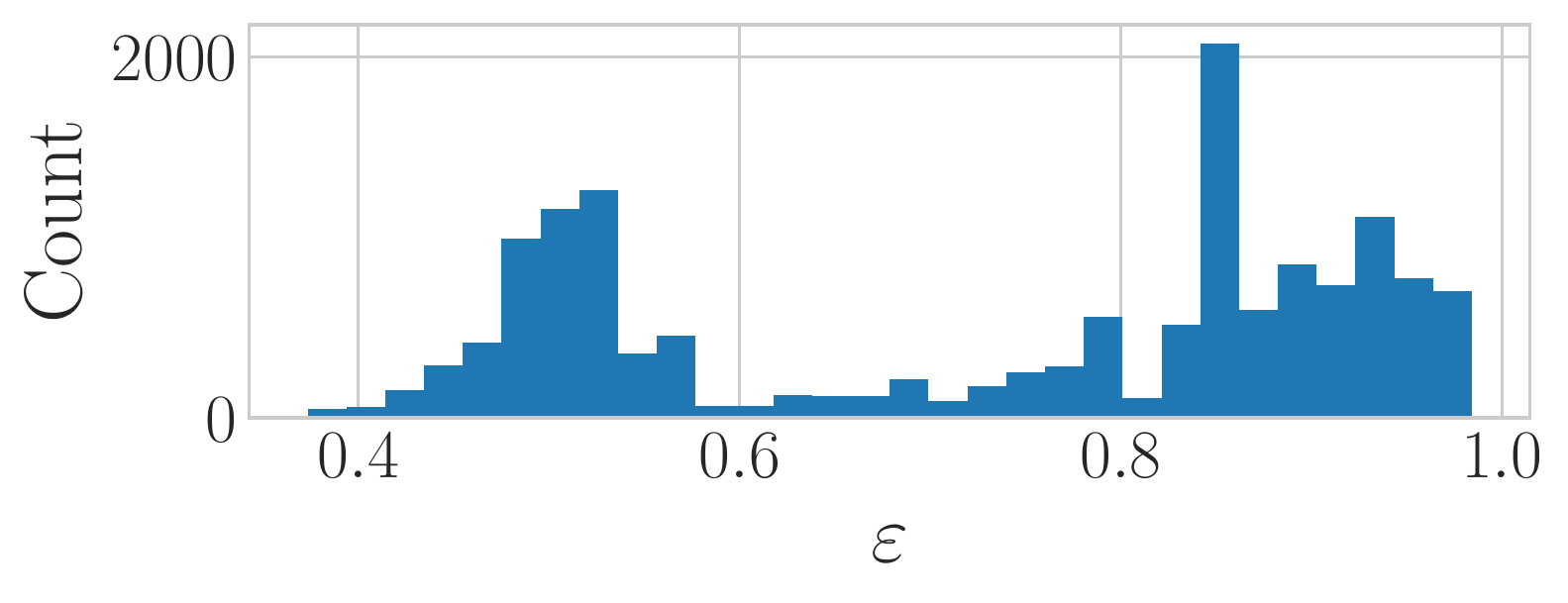}
    \caption{$\ell = 2000$}
    \label{fig: histogram_1000}
  \end{subfigure}%

  \begin{subfigure}[t]{0.24\textwidth}
    \centering
    \includegraphics[width = 1 \linewidth, trim={0.7em 0.7em 0.75em 0.7em}, clip]{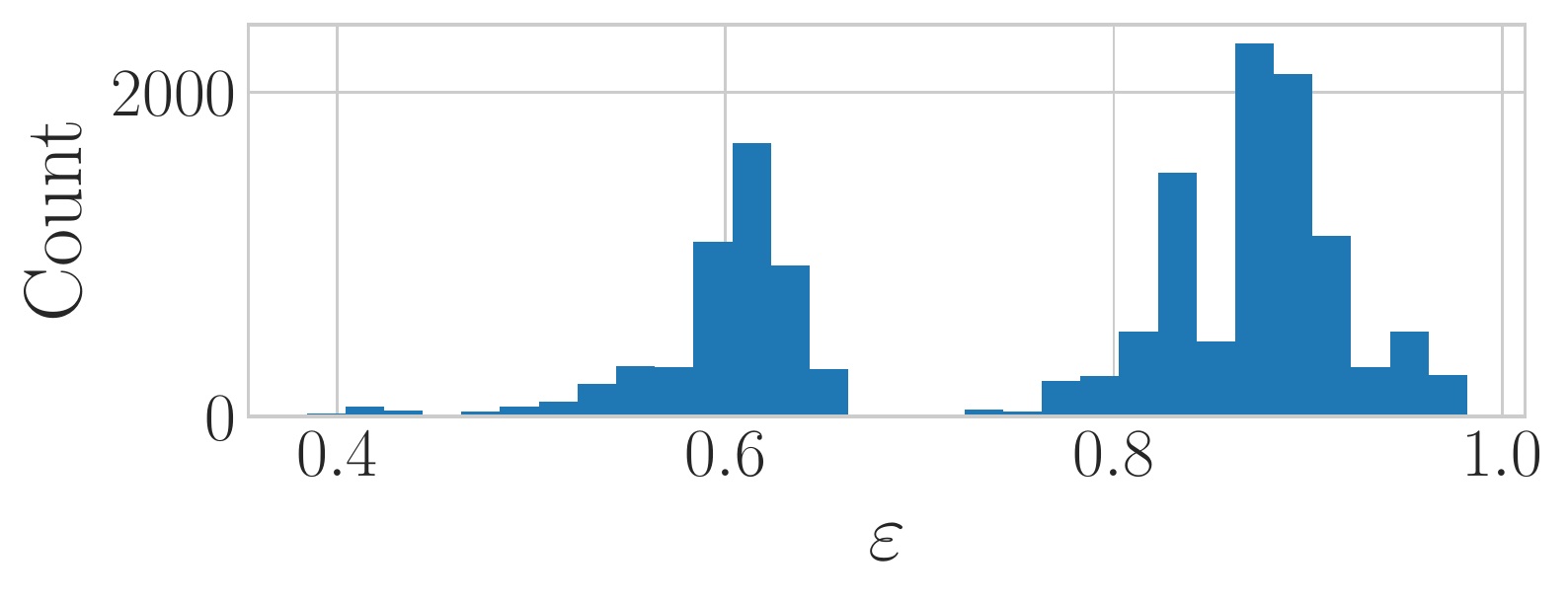}
    \caption{$\ell = 3000$}
    \label{fig: histogram_1500}
  \end{subfigure}%
  ~
  \begin{subfigure}[t]{0.24\textwidth}
    \centering
    \includegraphics[width = 1 \linewidth, trim={0.7em 0.7em 0.75em 0.7em}, clip]{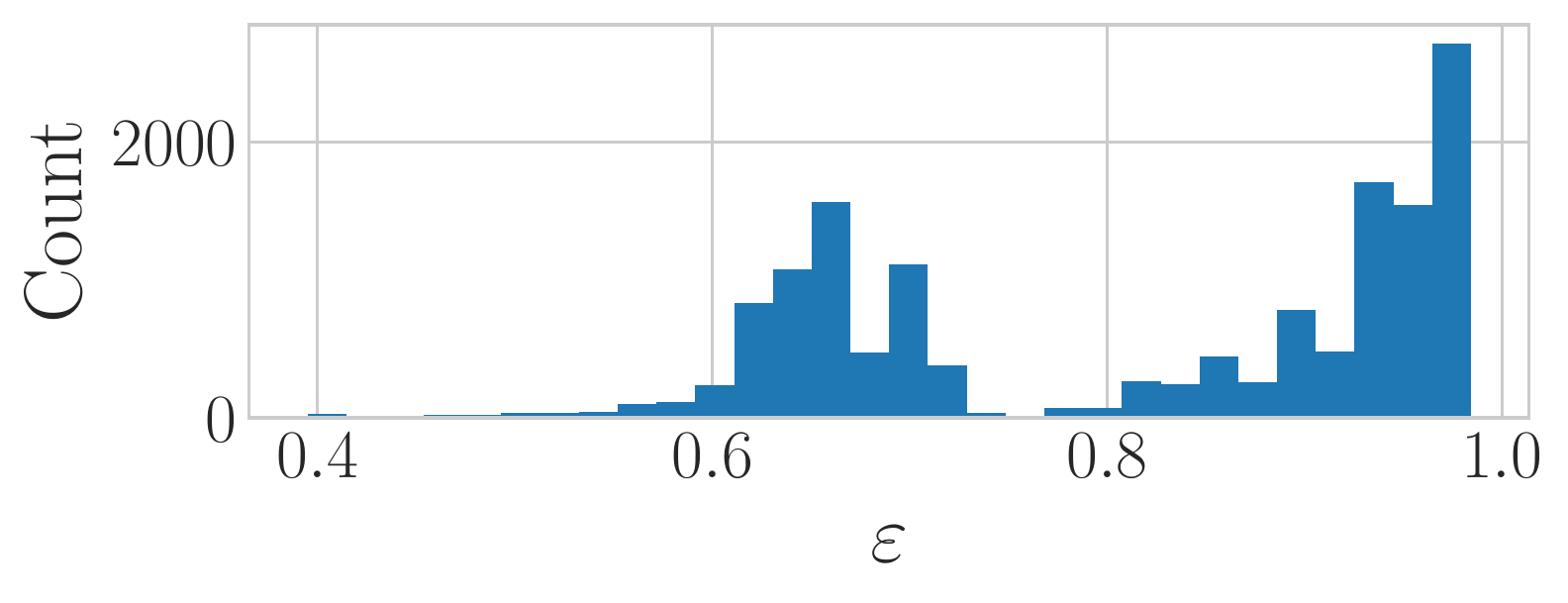}
    \caption{$\ell = 4000$}
    \label{fig: histogram_2000}
  \end{subfigure}%
  \caption{Histogram of per-agent $\varepsilon$ for varying privacy region edge length. We include all $32$ runs ($32$ (runs) $\times$ (17 + 154 + 116 + 174) (agents) = $14752$ data points).}
  \Description{Histogram of per-agent privacy loss.}
  \label{fig: histogram}
\end{figure}

Figure \ref{fig: histogram} plots the histogram of the per-agent $\varepsilon$ for varying privacy region edge length ($\ell$). For $\ell = 1000$ (Figure \ref{fig: histogram_500}), only $3572$ out of $14752$ agents ($24.2\%$) have $\varepsilon > 0.75$. This is because the majority of the agents converge fast~\cite{ijcai201931}, thus only a small percentage of them exhaust their budget. In fact, almost half of the total agents ($6759$ / $14752$, or $45.8\%$) have $\varepsilon \leq 0.5$. It is clear that the vast majority of agents benefit from really high degree of privacy.
% This is consistent with the findings of \cite{triastcyn2020data} that privacy loss is high for only a small portion of data/agents.

% trim={<left> <lower> <right> <upper>}

% % trim={<left> <lower> <right> <upper>}
% \begin{figure}[t!]
%     \centering
%     \includegraphics[width = 1 \linewidth, trim={0.7em 0.7em 0.75em 0.7em}, clip]{./Results/histogram_500.pdf}
%     \caption{Histogram of per-agent $\varepsilon$ for privacy region edge length $\ell = 4000$. We include all $32$ runs ($32$ (runs) $\times$ (17 + 154 + 116 + 174) (agents) = $14752$ data points).}
%     \Description{Histogram of per-agent privacy loss.}
%     \label{fig: histogram}
% \end{figure}

% trim={<left> <lower> <right> <upper>}

\subsection{Regions, Representative Agents, and Noise} \label{sec: Regions and Representative Agents}

In addition to the advantages of PLDP described in Section \ref{Advantages of PLDP}, there is another, more practical advantage that stems from the use of domain knowledge. The fragmentation function $\phi(\cdot)$ and the choice of the representative agent per region are domain specific. If the problem at hand (and by extension the utility function of the participating agents) is such that the representative agent has similar utilities to other agents in the region (and if we properly select the correct representative agent so that he is indicative of the agents in the region), then the social welfare will not degrade much, even under really strict budgets. Acting according to the representative agent, in such cases, allows for more informed allocations. This is a fundamental difference compared to, e.g., geo-indistinguishability, where the social welfare degrades in a significantly higher rate (as demonstrated in Figure \ref{fig: socialWelfare}). The latter can also be important for outlier agents, whose privacy cost per round might be high and thus lack the budget to play according to their own utilities for many rounds.

Regarding the choice of the fragmentation function $\phi(\cdot)$, there is a clear trade-off between the region size and the privacy cost per round, which in turn informs the amount of noise ($\zeta_S$ and $\zeta_B$). Restricting our privacy guarantees to a region helps reduce the required noise, since all the agents in a region have similar preferences (less noise is needed to become indistinguishable). If the privacy cost per round is small, an agent can afford lower noise (larger $\zeta_S$ and $\zeta_B$). Alternatively, acting according to the utilities of a properly chosen representative agent will still result in high quality allocations (especially in smaller regions, e.g., $\ell = 1000$), thus an agent might choose to accept higher noise in order to end up with much lower privacy cost at the end.

Finally, while in this work $\zeta_S$ and $\zeta_B$ are the same for all agents (see Section \ref{sec: Evaluation}), one can potentially achieve better results using adaptive noise. For example, agents can assume lower noise for the first few time-steps, and gradually increase it over time. Note, that the noise selection scheme must not depend on the agents' preferences. We leave this open for future work.

\section{Test-Case 2: Paper Assignment} \label{sec: Paper Assignment}

We ran a second test-case (Appendix \ref{supp: Paper Assignment}), where we use PLDP and PALMA to protect the reviewers' preferences during the paper assignment phase of a conference, using real data form \cite{10.1145/1458082.1458230}. PALMA achieved similar results: loss in social welfare $< 22\%$ (the maximally private solution loses $71.5\%$); $\varepsilon \leq 1$ and a median value of $0.36$. 
% We refer to the supplement for details.
% \footnote{It is harder to achieve high social welfare because there are almost 3 times more agents than resources, thus many agents end up without a match. The maximally private solution loses $71\%$}

\section{Conclusion} \label{sec: Conclusion}

Bridging the gap between physical and cyber worlds will bring about significant privacy risks and the potential to reveal highly sensitive information of users. In this paper, we consider the problem of hiding the utility function in multi-agent coordination problems. We propose \emph{PALMA}, a practical and scalable privacy-preserving algorithm for weighted matching along with PLDP, a `context-aware' privacy model that takes into account the `distance' between two utility functions. This ensures indistinguishability between agents with similar preferences. PALMA is decentralized, runs on-device, requires no inter-agent communication, converges in constant time under reasonable assumptions, and provides a \emph{strong level of privacy} ($\varepsilon \leq 1$ and median as low as $= 0.5$), while achieving high quality matchings (up to $86\%$ of the non-private optimal). To the best of our knowledge, we are the first to develop a practical and scalable framework for weighted matching and resource allocation in general, unboundedly large, multi-agent systems.

\subsection*{Acknowledgements}
This research was partially supported by TAILOR, a project funded by EU Horizon 2020 research and innovation programme under GA No 952215

% \section{Notes} \label{Notes}

% PALMA Can be used with traditional LDP. If radius is $\inf$, then PLDP $\rightarrow$ LDP.

% \note{We assume the graph G is a strongly connected graph, i.e., a path exists between any pair of vertices. \cite{prorok2017privacy}.\\ You would have to split it into regions, and you won't have privacy between regions?}

%%%%%%%%%%%%%%%%%%%%%%%%%%%%%%%%%%%%%%%%%%%%%%%%%%%%%%%%%%%%%%%%%%%%%%%%

%%% The next two lines define, first, the bibliography style to be 
%%% applied, and, second, the bibliography file to be used.
\clearpage
\balance
\bibliographystyle{ACM-Reference-Format} 
\bibliography{arxiv_palma_bibliography}

%%%%%%%%%%%%%%%%%%%%%%%%%%%%%%%%%%%%%%%%%%%%%%%%%%%%%%%%%%%%%%%%%%%%%%%%

\clearpage
\appendix
\section*{Appendix}

\subsection*{Contents}

In this supplementary material we include several details that have been omitted from the main text due to space limitations. In particular:

\begin{itemize}
	\item[-] In Section \ref{supp: Differential Privacy Definition} we explain the traditional Differential Privacy definition.
	\item[-] In Section \ref{supp: Renyi Divergence Definition} we provide the definition for the R\'enyi divergence.
	\item[-] In Section \ref{supp: PALMA supp} we provide additional implementation and complexity details on PALMA.
	\item[-] In Section \ref{supp: Mobility on Demand} we provide some additional details on the mobility-on-demand test-case.
	\item[-] In Section \ref{supp: Paper Assignment} we present the paper assignment test-case.
	\item[-] Finally, in Section \ref{sec: Societal Impact} we shortly discuss the societal impact.
\end{itemize}

% \item[-] Finally, in Section \ref{supp: Limitations} we shortly discuss the limitations of the proposed approach, and in Section \ref{sec: Societal Impact} the societal impact.

For narrative purposes, parts of the text of the main paper are repeated.

\section{Differential Privacy Definition} \label{supp: Differential Privacy Definition}

In this section we provide a short description of the traditional Differential Privacy (DP) definition;~\cite{dwork2006,dwork2006our,dwork2006calibrating} we refer the interested reader to~\cite{triastcyn2020data,dwork2014algorithmic} for a more comprehensive overview of Differential Privacy and Differential Privacy mechanisms.

Differential privacy is often discussed in the context of identifying individuals whose information may be in a database. It relies on an important impossibility result: impossibility of absolute disclosure prevention. The authors of~\cite{dwork2006,dwork2006our,dwork2006calibrating} prove that the conventional requirement of statistical database privacy -- access to a database should not allow an adversary to learn additional information about an individual than what could be learned without such access -- cannot be achieved due to \emph{auxiliary information} available to the adversary (besides the access to the database). As such, the authors argue to switch from absolute privacy guarantees to relative ones: informally, differential privacy captures the increased risk to an individual's privacy incurred by participating in a database. An algorithm is then considered differentially private if an adversary can not infer if a particular individual's information was used in the computation, given the output of said algorithm.

In order to achieve differential privacy, one needs a source of randomness. Let $\mathcal{M} : \mathcal{D} \rightarrow \mathcal{A}$ be a random function, mapping sensitive inputs from domain $\mathcal{D}$ to range $\mathcal{A}$ of privatized (or sanitized) outputs. In the context of matching problems in multi-agent systems, $\mathcal{D}$ can be the space of utility functions, and $\mathcal{A}$ the action space. Definition \ref{def:differential_privacy} defines a relaxation of differential privacy, called \emph{Approximate Differential Privacy} or \emph{$(\varepsilon, \delta)$-Differential Privacy}~\cite{dwork2014algorithmic}, which is more often used in artificial intelligence (and machine learning). 

\begin{definition}[$(\varepsilon, \delta)$-Differential Privacy]
\label{def:differential_privacy}
A randomized function (algorithm) $\mathcal{M}: \mathcal{D} \rightarrow \mathcal{A}$ with domain $\mathcal{D}$ and range $\mathcal{A}$ satisfies $(\varepsilon, \delta)$-differential privacy if for any two adjacent inputs $D, D' \in \mathcal{D}$ and for any set of outcomes $\mathcal{S} \subset \mathcal{A}$ the following holds:
\begin{align*}
  \Pr\left[\mathcal{M}(D) \in \mathcal{S} \right] \leq e^\varepsilon \Pr\left[\mathcal{M}(D') \in \mathcal{S} \right] + \delta.
\end{align*}
\end{definition}

\subsection{Intuitive Example}

In what follows, we provide some intuition on the interpretation of the $(\varepsilon, \delta)$ values (glossing over some of the technical details).

Imagine a simple, stripped-down example where there is only one agent $n$, and two resources $r_1$ and $r_2$. Suppose that agent $n$ prefers resource $r_1$, i.e., $u(r_1) > u(r_2)$. Under no regard for privacy, the optimal strategy for $n$ is to acquire resource $r_1$. However, an outsider observing his action will immediately know agent $n$'s preference. To protect privacy under DP, the agent will randomize its decisions by flipping a coin. Depending on the result (heads or tails), agent $n$ would acquire either resource $r_1$ or $r_2$, respectively. Now the observer can not know if the decision was taken based on the agent's actual preference, or due to the coin toss (plausible deniability). If the coin is unbiased it is easy to see that agent $n$'s preference is completely lost in the randomness and privacy is fully protected, but there is no utility benefit compared to a random allocation. \emph{This corresponds to $\varepsilon = 0$}. To increase the utility of the allocation, we will bias the coin towards the preferred resource $r_1$. Landing on heads is now more probable than landing on tails, and the ratio $Pr[\text{heads}] / Pr[\text{tails}]$ is greater than 1; \emph{$\varepsilon$ is the logarithm of this ratio}. The DP literature also refers to $\varepsilon$ as \emph{privacy budget}. Finally, imagine that sometimes the agent fails to flip a coin and just goes for the preferred resource. $\delta$ refers to this \emph{failure probability} (typically very small). In other words, an $(\varepsilon, \delta)$-Differentially Private algorithm provides a privacy guarantee $\varepsilon$ with probability $(1 - \delta)$. As such, the pair of these two values fully characterizes the privacy guarantee.

\section{R\'enyi Divergence Definition} \label{supp: Renyi Divergence Definition} 

The R\'enyi divergence of order $\lambda$ is defined as \cite{triastcyn2020data}:
\begin{align}
\label{def:renyi}
	\mathcal{D}_\lambda (P \| Q) &= \frac{1}{\lambda - 1} \log \mathbb{E}_p \left[ \left(\frac{p(x)}{q(x)}\right)^{\lambda-1} \right] dx  \\
		&= \frac{1}{\lambda - 1} \log \mathbb{E}_q \left[ \left(\frac{p(x)}{q(x)}\right)^{\lambda} \right] dx,
\end{align}
\noindent
where $\lambda$ is a hyper-parameter (assume for simplicity $\lambda \in \mathbb{N}$).

Analytic expressions for R\'enyi divergence exist for many common distributions and can be found in~\cite{gil2013renyi}. \cite{van2014renyi} provides a good survey of R\'enyi divergence properties in general.

Note that since our selection and back-off distributions are mixtures of two categorical distributions (see Equations \ref{Eq: selection probability} and \ref{Eq: backoff probability}), it is simple to compute the R\'enyi divergence.

\section{PALMA: A Privacy-Preserving Maximum-Weight Matching Heuristic} \label{supp: PALMA supp}

\subsection{Bounding the Set of Desirable Resources} \label{supp: Resource Set}

An important characteristic of many real-world applications is that there is typically a cost associated with acquiring a resource. As a result, each agent is typically interested in a subset of the total resources, i.e., $\mathcal{Q}^n \subset \mathcal{R}$. For example, a taxi driver would not be willing to drive to the other end of the city to pick up a low fare passenger, a driver would not be willing to charge his vehicle at a station in a different part of the city, and a reviewer would not be willing to review a paper outside his scope of expertise. This results in faster convergence (\emph{constant} time, see Section \ref{supp: Convergence}), but can also potentially lead to higher social welfare\footnote{The agent will loop back to $\mathcal{R}^n_1$, increases his chances of winning a high utility resources, instead of moving through a large number of undesirable resources.}. The sets ($\mathcal{R}^n_1, \dots, \mathcal{R}^n_i, \dots, \mathcal{R}^n_{R}$) can be contracted in the same manner as before.

% =====================================
% NOTE REGARDING mapBoundR
% DO NOT DELETE
% =====================================
% When we say an agent is interested in mapBoundR resources, it can not be that he has utility 0 for the rest. He has to have positive utility for the mapBoundR resources of its neighbors. So then he can play according to those probabilities. 
% Suppose that he plays with the non-zero probabilities, but then get rewards 0, this wouldn't work either, because the agent tries to acquire a resource thinking that it is a good one, while it is not.
% So in fact, in a realistic model, if we want mapBoundR = X, each agent would "declare" a mapBoundR << X, so if we include the union of all the agents in the neighborhood, we end up with the desired mapBoundR, and all agents in the neighborhood have positive utilities for those resources.

\subsection{Convergence} \label{supp: Convergence}

Theorem 2.1 of \cite{ijcai201931} proves that PALMA converges in polynomial time. In fact, under the aforementioned assumption that each agent is interested in a subset of the total resources (i.e., $\mathcal{Q}^n \subset \mathcal{R}$) and thus at each resource there is a bounded number of competing agents ($\mathcal{V}^r \subset \mathcal{N}$) Corollary 2.1.1 of \cite{ijcai201931} proves that the expected number of steps any individual agent requires to converge is independent of the total problem size (i.e., $N$ and $R$). In other words, by bounding these two quantities (i.e., we consider $|\mathcal{Q}^n|$, $|\mathcal{V}^r|$ to be constant functions of $N$, $R$), the convergence time is \emph{constant} in the total problem size $N$, $R$.

The initialization of PALMA is linear to the size of the region, $\mathcal{O}(\max_i|\mathcal{D}_i|)$, but this can be done once off-line. Finally, the accounting of the privacy loss is $\mathcal{O}(1)$.

\section{Computational Resources} \label{supp: Computational Resources}

All the simulations were run on a laptop equipped with an Intel i7-6820HQ CPU at 2.70GHz with 32.0 GB of RAM.

\begin{algorithm*}[!t]
\caption{Method for obfuscating the geo-location coordinates of agents and resources (based on~\cite{andres2013geo}).}
\label{tb: geo-ind algo}
\begin{tabular}{@{}l@{}}
% \toprule
Obfuscating geo-location $(lat, lon)$ by drawing a point $(r, \theta)$ from a polar Laplacian               \\ \midrule
1. Draw $\theta$ uniformly in $[0, 2\pi)$                           \\
2. Draw $p$ uniformly in $[0, 1)$ and set $r = C_\epsilon^{-1}(p)$, where $C_\epsilon^{-1}(p) = -\frac{1}{\epsilon} \left( W_{-1}(\frac{p - 1}{e}) + 1 \right)$, $\epsilon = \frac{\varepsilon}{l / 2}$, \\
$l$ is the privacy region's diameter, and $W_{-1}(\cdot)$ is the Lambert W function (the -1 brunch). \\
3. Set $dx = r \cos(\theta)$ and $dy = r \sin(\theta)$\\
4. Set $lat = lat + (dy \times 0.00000899)$ and $lon = lon + (dx \times 0.00000899) / \cos(lat \times \pi / 180)$, \\
where $0.00000899$ is one meter in degrees, calculated as 1 over the earth's radius in meters. \\ \bottomrule
\end{tabular}
\end{algorithm*}

\section{Test-Case 1: Mobility on Demand} \label{supp: Mobility on Demand}

\subsection{Setting} \label{supp: Mobility on Demand - Setting}

In the ridesharing scenario, we face repeated weighted matching problems; after a driver drops off a passenger, he is matched with a new one. Usually the matching process is performed in batches (e.g., every 10s). Assuming there is no vehicle relocation between the last drop off and the next match, we might have information leakage on the drop off location of the last passenger. To avoid this problem, we can use one-time ids for both the taxis and the passengers in every match, since both sets change dynamically anyway. Note that this problem is \emph{only relevant in this domain}; other applications, like the paper assignment problem, are not susceptible to this vulnerability.

\section{Test-Case 2: Paper Assignment} \label{supp: Paper Assignment}

\subsection{Setting} \label{supp: Paper Assignment - Setting}

In this test-case, we protect the reviewers' preferences during the paper assignment phase of a conference. We used the multi-aspect review assignment evaluation dataset \cite{paperAssignmentDataset}. It contains $73$ papers (which corresponds to the set of resources $\mathcal{R}$ in our setting) from the ACM SIGIR conference of 2007, and $189$ prospective reviewers (which corresponds to the set of agents $\mathcal{N}$) composed by authors of published papers in the top information retrieval conferences between 1971-2006. Each paper and each reviewer is represented by a $25$-dimensional binary label, representing one of the 25 major areas of ACM SIGIR \cite{10.1145/1458082.1458230}.

We used the $25$ major areas to define the privacy regions. Specifically, for each reviewer and paper, we selected uniformly at random one of the subject areas that they belong to, and set it as the \emph{primary} subject area. The primary subject area is unique, and identifies the region. The proposed Piecewise Local Differential Privacy demands that users belonging to the same region be indistinguishable from the attacker's point of view. This would correspond to reviewers with the same primary subject area. We refer to the remaining subject areas as \emph{secondary}. The maximum number of secondary subject areas of any adversary in a region defines the range of that region (reviewers are indistinguishable in that range). In this test-case, we consider adversaries with at most $2$, $3$, and $4$ additional subject areas\footnote{This would correspond to cosine distance of $\leq 0.2$, $\leq 0.25$, and $\leq 0.3$, respectively, from an agent that has a single subject area; the primary subject area of the corresponding region}. In layman's terms, a reviewer would be indistinguishable from any other reviewer that has the same primary subject area, and is an expert in at most $3$, $4$, and $5$ areas in total.

Finally, for each paper and reviewer, we convert the $25$-dimensional binary label to a continuous-valued vector. Specifically, the primary subject area is assigned the value $1$, all the secondary subject areas are assigned the value $0.5$, and the rest of the areas are assigned the value $0.1$. The latter reflects the fact that conferences trust the expertise of reviewers to asses the quality of papers in a broader area. Following the literature \cite{10.5555/3171642.3171649}, we used the cosine similarity (Equation \ref{eq: cosine}) of their label vectors to compute the utility of a paper to a reviewer.

\begin{equation} \label{eq: cosine}
	u_n(r) = \frac{\vec{n} \cdot \vec{r}}{\| \vec{n} \| \| \vec{r} \|}
\end{equation}

\noindent
where $\vec{n}$ ($\vec{r}$) denotes the $25$-dimensional label of agent $n$ (resource $r$).

Note that in a real-world paper assignment scenario, each reviewer would be required to review more than one paper (i.e., our matching graph would be a bipartite hypergraph). This can be easily handled by PALMA. Specifically, each reviewer will be represented by $x$ `copies', where $x$ is the number of papers each reviewer should review. Then, a resource (paper) would only signal agent $n$ that it is free (line 20 of Algorithm \ref{algo: palma}) if (i) it has been assigned to less than $y$ agents -- where $y$ represents the number of reviews per paper -- and (ii) a `copy' of agent $n$ has not acquired the resource. Nevertheless, this is out of the scope of this paper; the goal of this test-case is to provide additional evidence on the performance of PALMA on \emph{real data}. Thus, we opted to assign each reviewer to only one paper.

\subsection{Baselines}

As before, we employ the centralized Hungarian algorithm \cite{kuhn1955hungarian} to compute the non-private optimal -- in terms of social welfare -- solution, which we use to compute the loss in social welfare of PALMA. Note that in this test-case we do not compare to any geo- indistinguishability~\cite{andres2013geo} baselines because geo-indistinguishability is not directly applicable in this domain and modifying it to fit the domain is out of the scope of this paper. To the best of our knowledge, there is no other privacy preserving weighted matching algorithm to compare to.

We set $\zeta_S = 0.1$, $\zeta_B = \gamma = 0.05$, $B_n = 1$, $\lambda = 32$.

\subsection{Simulation Results: Social Welfare} \label{supp: Paper Assignment - Social Welfare}

% trim={<left> <lower> <right> <upper>}
\begin{figure}[t!]
    \centering
    \includegraphics[width = 1 \linewidth, trim={0.0em 0.5em 0.0em 0.5em}, clip]{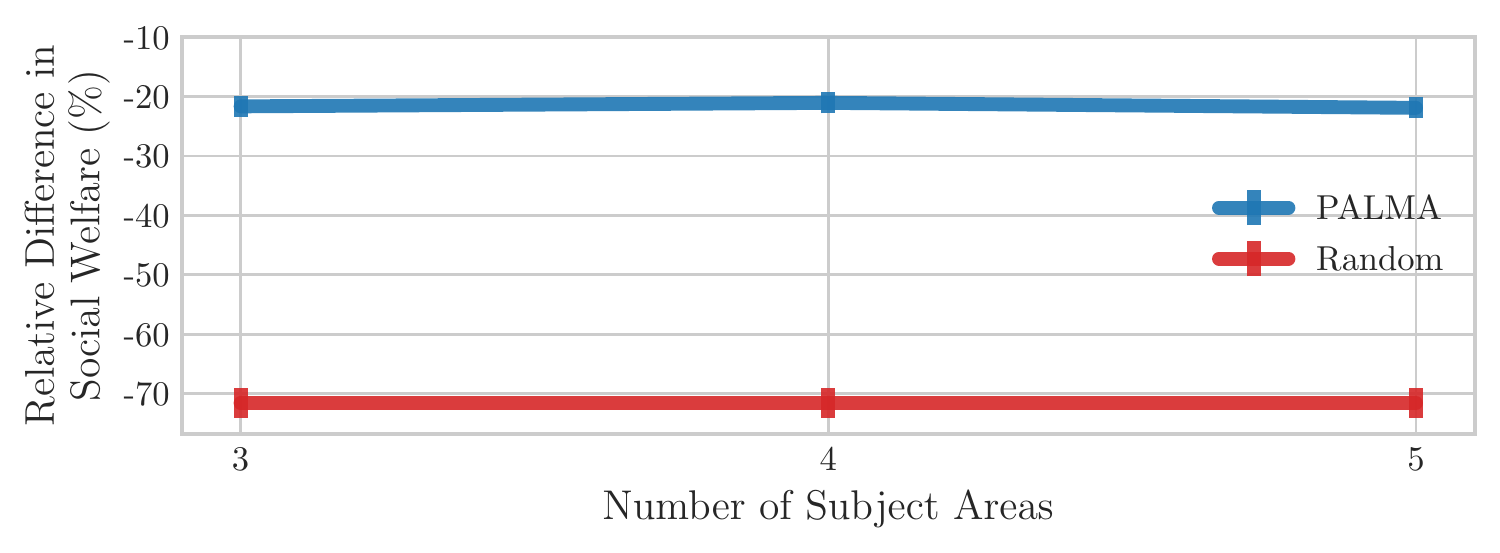}
    \caption{Loss in social welfare compared to the non-private, optimal solution for increasing size of the privacy region (i.e., number of subject areas).}
    \Description{Loss in social welfare compared to the non-private, optimal solution for increasing size of the privacy region (i.e., number of subject areas).}
    \label{supp fig: socialWelfare paper assignment}
\end{figure}

For $\varepsilon = 1$ given $\delta = 10^{-5}$ (Figure \ref{supp fig: socialWelfare paper assignment}), PALMA loses between $21.1 \pm 1.8\%$ to $21.9 \pm 1.8\%$ in social welfare compared to the non-private, optimal solution. The maximally private solution (i.e., the centralized random), losses $71.6 \pm 2.5\%$.

Contrary to the Mobility-on-Demand test-case, we observe a drop in social welfare. This is because in this test-case the number of agents is $2.6$ times bigger than the number of resources ($189$ reviewers vs. $73$ papers in the dataset). As a result, the majority of the reviewers remain un-matched. This does not constitute a problem for the centralized Hungarian, since it can compute a maximum-weight matching. Yet, in a randomized algorithm like PALMA, having an agent randomly back-off can lead to a drop in solution quality, as the majority of them will end up without a resource (i.e., zero reward). This is also reflected in the dramatic drop in social welfare of the random solution, which now losses $71.6\%$ compared to the $49.4\%$ loss in the Mobility-on-Demand test-case. This also suggests that in a real-world setting, where the number of papers is actually larger than the number of reviewers, PALMA will be able to close the gap in social welfare compared to the optimal solution. 

% Nevertheless, PALMA still significantly outperforms the other decentralized baseline, i.e., ALMA + geo-ind.

\subsection{Simulation Results: Privacy} \label{supp: Paper Assignment - Privacy Loss}

% trim={<left> <lower> <right> <upper>}
\begin{figure}[t!]
    \centering
    \includegraphics[width = 1 \linewidth, trim={0.0em 0.5em 0.0em 0.5em}, clip]{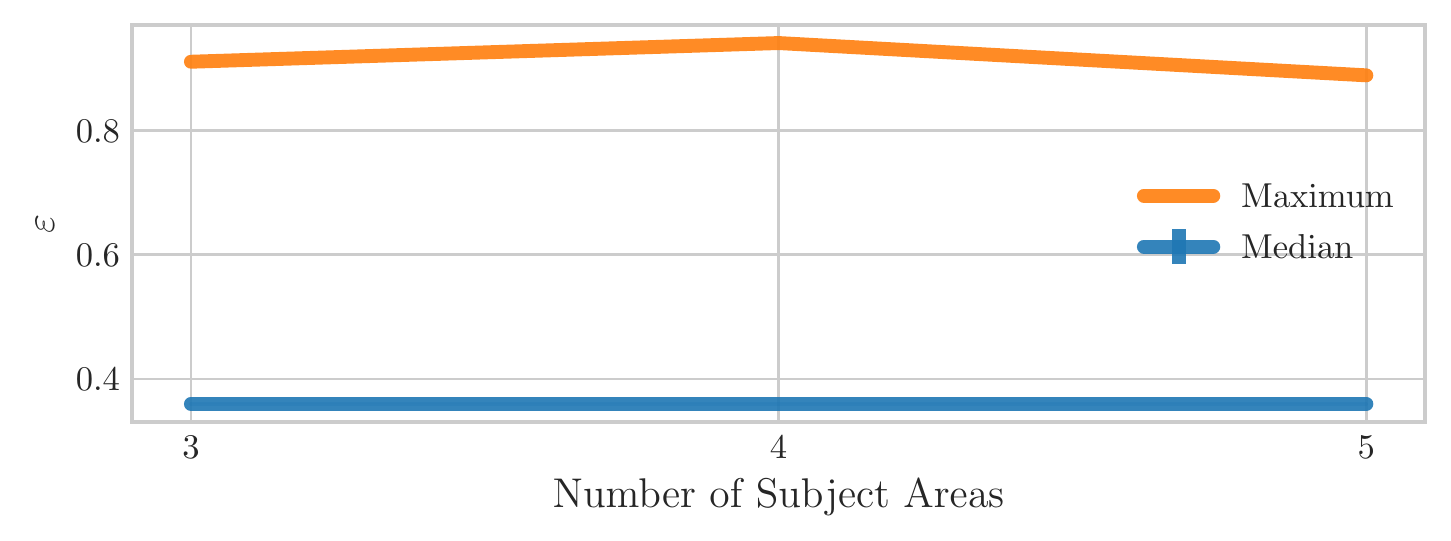}
    \caption{Maximum (orange line) and median (blue line) per-agent $\varepsilon$ for increasing values of the size of the privacy region (i.e., number of subject areas).}
    %  The shaded area represents the range between the maximum and minimum value of the median.
    \Description{Maximum (orange line) and median (blue line) per-agent privacy loss for increasing values of the size of the privacy region (i.e., number of subject areas).}
    \label{supp fig: privacyLoss paper assignment}
\end{figure}

% % trim={<left> <lower> <right> <upper>}
% \begin{figure}[t!]
%     \centering
%     \includegraphics[width = 1 \linewidth, trim={0.0em 0.5em 0.0em 0.5em}, clip]{./Results/histogram_paperAssignment_2.pdf}
%     \caption{Histogram of per-agent $\varepsilon$ for a privacy region with 3, 4, or 5 subject areas. We include all 32 runs (32 (runs) $\times$ 73 (agents matched with a resource) $\times$ 3 (test-cases with different number of subject areas) = 7008 data points in total).}
%     \Description{Histogram of per-agent privacy loss.}
%     \label{supp fig: histogram paper assignment}
% \end{figure}

Figure \ref{supp fig: privacyLoss paper assignment} depicts the maximum (out of all the 32 runs) and median (average median value over the 32 runs) per-agent $\varepsilon$ for increasing values of the size of the privacy region (i.e., number of additional subject areas). The average value of the median is $0.36$. Only between $0.9 - 2.1\%$ of the agents have $\varepsilon > 0.75$ (for the three privacy regions cases). The maximum per-agent $\varepsilon$ is bounded by the privacy budget (i.e., $\varepsilon = 1$).

% Figure \ref{supp fig: histogram paper assignment} plots the histogram of the per-agent $\varepsilon$ for a privacy region with 3 subject areas. Only $1150$ out of $7008$ agents ($16.4\%$) have $\varepsilon > 1$. This is because the majority of the agents converge fast~\cite{ijcai201931}, thus only a small percentage of them exhaust their budget. In fact, more than half of the total agents ($4740$ / $7008$, or $67.6\%$) have $\varepsilon \leq 0.5$. It is clear that the vast majority of agents benefit from really high degree of privacy.

% \section{Limitations} \label{supp: Limitations}

% In the extreme scenario where an adversary has enough background knowledge to narrow down his search to two agents that belong to different neighborhoods then, if he can figure out the neighborhood of his target, he can figure out the targeted agent. Yet, this scenario is unlikely, and it is more of a question on how to design regions in a way that the aforementioned problems are avoided.

% Moreover, one limitation is that we need to design regions.

\section{Societal Impact} \label{sec: Societal Impact}

The rapid proliferation of intelligent systems and autonomous agents has the potential to positively impact many facets of our daily lives. However, harnessing their power requires massive amounts of personal data to be collected, stored, processed, and analyzed -- often by resource-constrained devices. The latter has raised serious privacy concerns and has resulted in an accelerated growth of privacy advocacy movements. Our work shows that harnessing the potential of intelligent systems does not have to compromise privacy.

% We provide a \emph{practical} and \emph{applicable} framework -- PALMA can run \emph{on-device} -- for solving one of the fundamental problems of multi-agent systems (finding matches, and allocations), while providing \emph{strong} worst-case \emph{privacy} guarantees.

\clearpage

%%%%%%%%%%%%%%%%%%%%%%%%%%%%%%%%%%%%%%%%%%%%%%%%%%%%%%%%%%%%%%%%%%%%%%%%

\end{document}